\documentclass{article}
\usepackage{hyperref}

\usepackage{indentfirst}
\usepackage[margin=1in]{geometry}

\usepackage[utf8]{inputenc} 
\usepackage[T1]{fontenc}    
\usepackage{hyperref}       
\usepackage{url}            
\usepackage{booktabs}       
\usepackage{amsfonts}       
\usepackage{nicefrac}       
\usepackage{microtype}      
\usepackage{lipsum}		
\usepackage{graphicx}
\usepackage[numbers]{natbib}
\usepackage{doi}
\usepackage{mathtools}
\usepackage{amsmath}
\usepackage{tabularx,ragged2e,booktabs,caption}
\usepackage{tikz}
\usetikzlibrary{positioning}
\usepackage[makeroom]{cancel}
\usetikzlibrary{shapes,arrows}
\usepackage{caption}
\newcommand*{\h}{\hspace{5pt}}
\newcommand*{\hh}{\h\h}
\usetikzlibrary{matrix,shapes,arrows,positioning,chains}

\tikzset{
decision/.style={
    ellipse,
    draw,
    text width=10em,
    text badly centered,
    inner sep=3pt
},
block/.style={
    rectangle,
    draw,
    text width=14em,
    text centered,
    rounded corners
},
cloud/.style={
    draw,
    ellipse,
    minimum height=2em
},
descr/.style={
    fill=white,
    inner sep=2.5pt
},
connector/.style={
    -latex,
    font=\scriptsize
},
rectangle connector/.style={
    connector,
    to path={(\tikztostart) -- ++(#1,0pt) \tikztonodes |- (\tikztotarget) },
    pos=0.5
},
rectangle connector/.default=-2cm,
straight connector/.style={
    connector,
    to path=--(\tikztotarget) \tikztonodes
}
}

\makeatletter
\newcommand*{\centernot}{%
  \mathpalette\@centernot
}
\def\@centernot#1#2{%
  \mathrel{%
    \rlap{%
      \settowidth\dimen@{$\m@th#1{#2}$}%
      \kern.5\dimen@
      \settowidth\dimen@{$\m@th#1=$}%
      \kern-.5\dimen@
      $\m@th#1\not$%
    }%
    {#2}%
  }%
}
\makeatother

\newcommand{\independent}{\perp\mkern-9.5mu\perp}
\newcommand{\notindependent}{\centernot{\independent}}

\date{} 					
\begin{document}
\title{Frameworks for Estimating Causal Effects in Observational Settings: Comparing Confounder Adjustment and Instrumental Variables}

\author{{Roy S. Zawadzki}\thanks{Corresponding Author: zawadzkr@uci.edu}\\
	Department of Statistics\\
	University of California, Irvine\\
	Irvine, CA\\
	\and
	{Joshua D. Grill}\\
	Department of Psychiatry and Human Behavior\\
	Department of Neurobiology and Behavior\\
	University of California, Irvine\\
	Irvine, CA\\
	\and
	{Daniel L. Gillen}\\
	Department of Statistics\\
	University of California, Irvine\\
	Irvine, CA\\
	\and
	for the Alzheimer’s Disease Neuroimaging Initiative\thanks{Data used in preparation of this article were obtained from the Alzheimer’s Disease
Neuroimaging Initiative (ADNI) database (adni.loni.usc.edu). As such, the investigators
within the ADNI contributed to the design and implementation of ADNI and/or provided data
but did not participate in analysis or writing of this report. A complete listing of ADNI
investigators can be found at:
\href{http://adni.loni.usc.edu/wp-content/uploads/how_to_apply/ADNI_Acknowledgement_List.pdf}{http://adni.loni.usc.edu/wp-content/uploads/how\_to\_apply/ADNI\_Acknowledgement\_List.pdf}}
}

\hypersetup{
pdftitle={Principles for Estimating Causal Effects in Observational Settings},
pdfsubject={stat.ME},
pdfauthor={Roy S. Zawadzki, Joshua D. Grill, Daniel L. Gillen},
pdfkeywords={Causal inference,Observational studies,Instrumental variables,Adjustment,Propensity scores},
}

\maketitle

\begin{abstract}
	To estimate causal effects, analysts performing observational studies in health settings utilize several strategies to mitigate bias due to confounding by indication. There are two broad classes of approaches for these purposes: use of confounders and instrumental variables (IVs). Because such approaches are largely characterized by untestable assumptions, analysts must operate under an indefinite paradigm that these methods will work imperfectly. In this tutorial, we formalize a set of general principles and heuristics for estimating causal effects in the two approaches when the assumptions are potentially violated. This crucially requires reframing the process of observational studies as hypothesizing potential scenarios where the estimates from one approach are less inconsistent than the other. While most of our discussion of methodology centers around the linear setting, we touch upon complexities in non-linear settings and flexible procedures such as target minimum loss-based estimation (TMLE) and double machine learning (DML). To demonstrate the application of our principles, we investigate the use of donepezil off-label for mild cognitive impairment (MCI). We compare and contrast results from confounder and IV methods, traditional and flexible, within our analysis and to a similar observational study and clinical trial.
\end{abstract}

\section{Introduction}

A common goal in healthcare is to estimate the causal effect of a treatment or intervention on medical outcomes. The gold standard for this task remains a well-controlled randomized control trial (RCT). In this setting, randomization allows us to establish cause and effect estimates because, on average, observed differences between treatment arms will be due to either the assigned treatment or chance.

RCTs may not always be feasible due to ethical, logistical, or monetary constraints. When this is the case, we may turn to observational studies in an attempt to isolate causal effects of interest. Observational studies can provide hypothesis generating evidence to help inform future investigations of treatments such as extensions to different populations or use cases. In addition, observational studies give researchers real world evidence surrounding the effectiveness and safety of a treatment to augment RCT findings (e.g. Phase 4 trials).

On the path to isolating causal effects, observational studies must address potential bias due to the non-randomization of the intervention. A common source of bias is confounding by indication, or treatment selection bias, where factors affect both the assignment of treatment and the targeted medical condition. These factors, called confounders, range from patient characteristics to other concurrent treatments. 

There are two broad classes of approaches to mitigate treatment selection bias based on confounding variables and instrumental variables (IVs). Briefly, confounder approaches aim to "adjust" for all factors that both explain treatment assignment and the outcome. In contrast, IVs determine only the assignment of the treatment but, otherwise, are not associated with the outcome. IVs are used to define a subset of the population whose treatment assignment is free from confounding.

Fundamentally, confounder and IV approaches are characterized by untestable assumptions in practice. For example, the notion that all possible confounders have been adjusted for cannot be verified with data. Therefore, analysts must be able to operate under the assumption that these methods will work imperfectly. In other words, neither approach will fully overcome treatment selection bias but can provide a less biased estimate than had they not been used. Navigating this indefinite paradigm requires a general set of reasoning and intuition surrounding observational studies. 

In this paper, we formalize a set of general principles and heuristics for estimating causal effects under treatment selection bias. Stemming from the two approaches, we outline three general steps. Firstly, one must be able to identify potential confounders and IVs in the scientific context of the study. Then, looking at the available dataset, judge which variables identified in step one are present, somewhat present (e.g. proxies), or missing. Thirdly, weighing the pros and cons of each methodology to consider which one could provide a more reasonable causal estimate. Our discussion is led by the criteria of internal validity (the ability for the approach to estimate the causal effect), external validity (the ability for the approach to generalize to relevant populations), and reproducibility (the ability for the results to be replicated in similar studies). 

Throughout the paper, we will generally assume that the relationships in the data are linear in functional form so we may focus on issues related to unobserved confounding as opposed to model misspecification. We relax this linearity assumption when we speak about causal inference in non-linear settings and flexible modeling approaches.

To demonstrate the use of the principles outlined in the paper, as an illustrative example, we investigate the use of donepezil (brand name: Aricept) in mild cognitively impaired patients (MCI) to mitigate cognitive decline due to dementia. In the present day, donepezil is indicated for mild to severe Alzheimer's Disease (AD) but not MCI due to failure to show efficacy in clinical trials.\citep{petersen_vitamin_2005,salloway_efficacy_2004,doody_donepezil_2009} Nevertheless, study of donepezil and related compounds in MCI has continued in both observational studies of off-label practice \citep{sokolow_deleterious_2017,schneider_treatment_2011} and clinical trials.\citep{devanand_donepezil_2018,montero-odasso_donepezil_2019}

We analyze data from the Alzheimer's Disease Neuroimaging Initiate (ADNI), an observational, multi-center, natural history dataset that tracks cognitively normal, MCI, and AD subjects over time.\citep{petersen_alzheimers_2010} Because the use of donepezil is non-randomized in ADNI there may be confounding by indication where MCI patients who are prescribed donepezil are potentially suffering from more severe cognitive decline than those who do not. Under this setting, we will compare confounder and IV model estimates to each other and to that of related observational studies and clinical trials.

In the literature, there exists other tutorials and textbooks for causal inference methodology. In particular, Baiocchi, Cheng, and Small give a detailed overview of IVs in health research and, in section two of their paper, share a similar philosophy to this paper in that choosing between confounder and IV approaches includes weighing unmeasured confounding against the hypothesized validity of the IV.\citep{baiocchi_tutorial_2014} Nevertheless, the scope of their tutorial surrounds a pedological overview of IVs whereas ours incorporates this information as part of a larger goal to provide detailed guidelines and heuristics for conducting observational studies. As such, we cover a broader range of topics such as the interaction between the confounders and IV approaches, comparing confounder adjustment and propensity score weighting methods, non-collapsibility, and external validity with each topic centered around potential assumption violations. We further include a full data analysis with a detailed comparison and discussion of each approach in order to clearly demonstrate the discussed principles.

Another important contribution of our tutorial is that we cover the recent trend of using machine learning (ML) for causal inference such as targeted minimum loss-based estimation (TMLE) and double machine learning (DML).\citep{van_der_laan_targeted_2010,schuler_targeted_2017,chernozhukov_doubledebiased_2018} As these tools are now widely used in applied data analyses, any modern tutorial on causal inference methodology should include commentary on this topic. By first outlining general principles for causal inference in observational studies with “traditional” methodology, our tutorial can more effectively discuss the potential benefits and pitfalls of “data-driven” modeling for estimating causal effects. Furthermore, our applied data analysis, that includes the use of ML methods, provides a further critical evaluation of the novel ML methods. There exist little by way of practical comparisons and guidelines of traditional and novel methodology across confounder and IV methodology in the literature. Angrist and Frandsen consider this topic but focus on economic applications and the work is limited in terms of methodology utilized relative to this review.\citep{angrist_machine_2022}

The remainder of the paper is organized as follows. We begin by outlining the assumptions related to confounder adjustment and IV methods as well as the consequences of violating each assumption. We then describe common methodologies for confounder adjustment and IVs and their machine learning extensions, discussing further assumptions and considerations. Next, we outline scenarios where one approach is preferable to another. After presenting our analysis of the ADNI donepezil data and related remarks, we conclude with an overall discussion of the content of this paper.

\section{Two Approaches: Confounders and Instruments}
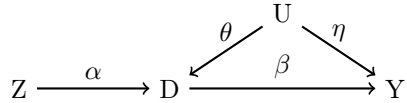
\begin{figure}[h]
    \centering
    \begin{tikzpicture}[thick]
        \node (1) at (0,0) {Z};
        \node (2) at (2,0) {D};
        \node (3) at (5,0) {Y};
        \node (5) at (3.5,1) {U};

        \path [->] (1) edge node[above]{$\alpha$} (2);
        \path [->] (2) edge node[above]{$\beta$} (3);
        \path [->] (5) edge node[above]{$\theta$} (2);
        \path [->] (5) edge node[above]{$\eta$} (3);
    \end{tikzpicture}
    \caption{A Directed Acyclic Graph with One Confounder and One IV}
    \label{fig:dag_simple}

\end{figure}

We begin by considering the simple scenario depicted in Figure \ref{fig:dag_simple}. Such figures are called directed acyclic graphs (DAGs) where the nodes are variables, the edges represent a directed causal effect, and the greek letters represent the magnitude of the causal effect of the respective edge. Contextually, $D$ is an indicator for which treatment was prescribed, $Y$ represents the outcome of interest, $U$ represents a confounding variable, and $Z$ is an IV. In the donepezil example, $D$ indicates the prescription of donepezil or not and $Y$ is cognitive function as measured by the Alzheimer's Disease Assessment Scale (ADAS-cog).\citep{kueper_alzheimers_2018}

Firstly, we must define the causal estimand of interest using the potential outcomes framework.\citep{rubin_causal_2005} Let $Y_{i,t}(1)$ be the potential outcome at time $t$ had the individual taken the treatment ($D = 1$) and $Y_{i,t}(0)$ be the potential outcome after 2 years had the individual not taken the treatment ($D = 0$). In contrast, $Y_{t}$ is the observed outcome at time $t$. Therefore, in the situation characterized by Figure \ref{fig:dag_simple}, we want to isolate $\beta$, the average treatment effect (ATE). In the donepezil example, the ATE would be change from baseline in ADAS-cog after two years formalized below with $t=0$ being baseline:

\begin{align*}
    \beta &= E\big[Y_{t=2}(1) - Y_{t=2}(0)\big] = E\big[Y_{t=2}(1) - Y_{t=2}(0) + (Y_{t=0} - Y_{t=0})\big]\\
    & = E\big[(Y_{t=2}(1) - Y_{t=0}) - (Y_{t=2}(0) - Y_{t=0})\big].
\end{align*}

We cannot observe both potential outcomes for any individual and thus we must use observed values from those who were prescribed either treatment option. Under the stable unit treatment values assumption (SUTVA), treatment assignment ignorability (i.e. $Y(0),Y(1) \independent D$), and positivity ($0 < P(D=d) < 1$ with $d=0,1$), we could use the following estimand

\begin{align*}
    \beta &= E\big[(Y_{t=2}(1) - Y_{t=0}) - (Y_{t=2}(0) - Y_{t=0})\big]\\
    &= E\big[(Y_{t=2}(1) - Y_{t=0})|D=1\big] - E\big[(Y_{t=2}(0) - Y_{t=0})|D=1\big]\\
    &= E[Y_{t=2} - Y_{t=0}|D=1] - E[Y_{t=2} - Y_{t=0}|D=0].
\end{align*}

Throughout this paper, we focus on mitigating issues when the assumption of ignorability fails to hold. Graphically, the $\theta$ and $\eta$ edge weights in Figure \ref{fig:dag_simple} represent the magnitude of $U$'s influence on the assignment of donepezil and ADAS-cog, respectively. Clearly we have a violation because $U$ is a confounder when $\theta \ne 0$ and $\eta \ne 0$ resulting in $Y(0),Y(1) \notindependent D$. Simply estimating the difference in the group means (the final lines) will not recover $\beta$ due to the second equality failing. In short, indirect paths from $D$ to $Y$ (via $U$) pose major issues in obtaining causal estimates.

There are two ways we will describe how the confounder approach can isolate $\beta$. First, we need to find a set of variables $\textbf{X}$ such that conditional ignorability, $D \independent Y(0),Y(1)|\textbf{X}$ is achieved. Alternatively, we need to condition upon variables $\textbf{X}$ such that all "backdoor" paths from $D$ to $Y$ are blocked.\citep{pearl_causal_1995} Backdoor paths are non-causal, indirect paths that connect $D$ and $Y$. Visually, if we think of a DAG as a set of pipes, blocking all backdoor paths restricts the flow of water or "information" to only go through the pipe flowing from $D$ to $Y$. Hence, we recover our direct causal effect. These two notions of confounding leads us to conclude that in Figure \ref{fig:dag_simple} it is sufficient and necessary to condition on $U$ and use the finite estimator of $E\big[E[Y_{t=2} - Y_{t=0}|D=1,U=u] - E[Y_{t=2}-Y_{t=0}|D=0,U=u]\big]$ to obtain $\beta$. Note that there are many other descriptions of confounding and we will focus on the conditional ignorability and graphical definitions. \citep{vanderweele_definition_2013}\citep{greenland_confounding_1999}

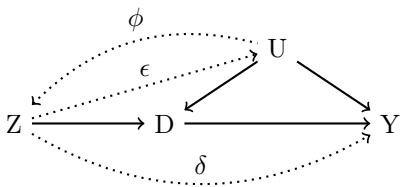
\begin{figure}[h]
    \centering
    \begin{tikzpicture}[thick]
        \node (1) at (0,0) {Z};
        \node (2) at (2,0) {D};
        \node (3) at (5,0) {Y};
        \node (5) at (3.5,1) {U};

        \path [->] (1) edge (2);
        \path [->] (1) edge[dotted,bend right=30] node[above] {$\delta$} (3);
        \path [->] (1) edge[dotted] node[above]{$\epsilon$}  (5);
        \path [<-] (1) edge[dotted,bend left=30] node[above]{$\phi$} (5);
        \path [->] (2) edge (3);
        \path [->] (5) edge (2);
        \path [->] (5) edge (3);
    \end{tikzpicture}
    \caption{Dotted lines disqualify $Z$ from being an IV}
    \label{fig:dag_z_violation}
\end{figure}

Alternatively, we may identify the treatment effect through an IV denoted as $Z$. For simplicity, we assume for now that only one IV is sufficient. Briefly, the definition of an IV is that $Z$ must influence $D$ (relevance), or $\alpha \ne 0$, does not cause $Y$ conditioning on $X$ (exclusion restriction), and is not associated with any unobserved confounders (independence). In Figure \ref{fig:dag_simple}, $Z$ is an IV because $\alpha \ne 0$ and there are no other arrows in or out of $Z$ that go to $Y$. Figure \ref{fig:dag_z_violation} demonstrates how this latter notion can be violated if either $\delta$, $\epsilon$, or $\phi$ are non-zero. In this case, $Z$ is, in fact, a confounder but if $\delta = 0$ and we are able to condition upon $U$, then $Z$ is re-classified as an IV. Compared with the confounder approach, if we have access to an IV, we may be able to obtain a causal estimate without having to account for all possible confounders, which is a major potential advantage of using IVs over confounder adjustment methods.

The building of a valid causal network requires both knowledge of all variables in the network and the arrows between them. Unfortunately, we cannot be sure that a posited DAG is correct using data. For example, to suggest unobserved confounding requires knowledge that goes beyond the dataset at hand. Nevertheless, we can still use DAGs to capture assumption violations as clearly as possible and move towards the best option in a given scenario. By first thinking outside the scope of the current dataset, one captures a fuller picture of the study.

The untestability of causal assumptions suggests that users of the confounder and IV approaches must think relatively and not in absolutes. For example, rather than arguing a set of confounders is sufficient for conditional ignorability, one should instead find confounders to condition upon that potentially bring the estimate closer to the true causal estimand. For IVs, rather than justifying whether have a true IV or not, we can think about how strongly the treatment is identified relative to potential violations in the assumptions of $Z$ and the hypothesized overall magnitude of confounding.

\subsection{Identifying and Using Confounders}

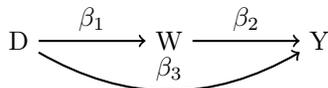
\begin{figure}[h]
    \centering
    \begin{tikzpicture}[thick]
        \node (1) at (0,0) {D};
        \node (2) at (2,0) {W};
        \node (3) at (4,0) {Y};
        
        \path [->] (1) edge node[above]{$\beta_1$} (2);
        \path [->] (2) edge node[above]{$\beta_2$} (3);
        \path [->] (1) edge[bend right=30] node[above] {$\beta_3$} (3);
    \end{tikzpicture}
    \caption{DAG of a Mediator, $W$}
    \label{fig:dag_mediator}
\end{figure}

Without directly modeling the response, in order to avoid multiplicity bias, a reasonable strategy to identify confounders is to first postulate variables that affect the response and then distinguish which of these variables may influence treatment assignment.\citep{brookhart_confounding_2010} In the latter step, caution must be taken in the direction of causality: mistakenly adjusting for mediating variables on the pathway from $D$ to $Y$ may produce unintended consequences such as attenuation of the estimated treatment effect. To see this, consider Figure \ref{fig:dag_mediator} – a scenario where $W$ is a mediating variable. A simple example of a meditator is the ADAS-cog measurement at one year, $Y_{t=1}$. Adjusting for $Y_{t=1}$ will decrease a possible treatment effect because the measurement at one year is on the causal pathway to $Y_{t=2}$ and we have blocked this path.

For a more formal attenuation scenario, denote the edge weights in Figure \ref{fig:dag_mediator} as $\{\beta_{i}: i = 1,2,3\}$ (the decomposed $\beta$ in Figure 1). If we assume linear relationships between variables then $\beta = \beta_1\beta_2 + \beta_3$. If $|\beta| > 0$ and $sign(\beta_1\beta_2))=sign(\beta_3)$, then adjusting for $W$ will yield $\beta_3$ but there will be attenuation as $|\beta|>|\beta_3|$.

The first and strongest reference for determining causal links for confounders should be the underlying scientific mechanism. Such information can be based on prior basic science, epidemiological findings, and historical trials. These sources often help one to identify a vast majority of relevant confounding factors. Another source but of potentially lesser quality is past empirical studies done on predictors of the response. One should assess the quality of these studies in terms of replicability, precision, and study design before choosing to use the associated information. This information can additionally be used to identify suitable proxies for key confounders.

Given a set of potential confounders, it may not always be advantageous to select all of them in the data analysis. In reality, much of the confounding may be captured by a few variables such as basic demographics (e.g. age and sex), commonly collected lifestyle factors (e.g. smoking and alcohol use), and comorbidities (e.g. chronic disease and corresponding medication use). With each confounder included in the analysis, we must weigh moving towards conditional ignorability against overfitting (i.e. increased imprecision and Type II error), interpretability, and reproducibility. Consider that under non-linearity, adjusting for confounders may change the interpretation of the estimate of the treatment effect.\citep{greenland_confounding_1999} In the linear setting, a similar scenario can occur under treatment effect heterogeneity, which is when the effect of the treatment differs across the values of one or more factors.\citep{aronow_does_2016}

\subsection{Identifying and Using Instrumental Variables}

\begin{figure}[h]
    \centering
    \begin{tikzpicture}[thick]
        \node (1) at (0,0) {A};
        \node (2) at (2,0) {D};
        \node (3) at (5,0) {Y};
        \node (4) at (1,-0.5) {Z};
        \node (5) at (3.5,1) {\xcancel{U}};
        
        \path [->] (1) edge (2);
        \path [->] (1) edge[dotted] (5);
        \path [<-] (1) edge[dotted,bend left=30] (5);
        \path [->] (1) edge (4);
        \path [->] (2) edge (3);
        \path [->] (5) edge (2);
        \path [->] (5) edge (3);
    \end{tikzpicture}
    \caption{$Z$ is a valid IV because we have blocked the path via $U$ and it is a proxy of an effectively randomized variable $A$. By varying $Z$, we allow variation in $A$ to influence $D$.}
    \label{fig:cond_valid_z}
\end{figure}
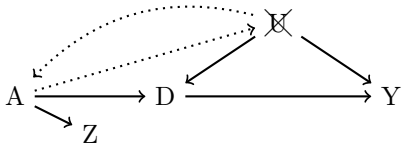

Unlike confounders, the use of IVs is not as straightforward and often requires more technical knowledge to employ effectively so we first will provide more background and intuition. The crux of the IV approach is that we use variation independent from confounding to identify treatment assignment. Because IVs, by definition, cannot be determined by other variables in the causal paradigm (there are exceptions: for example, see Figure \ref{fig:cond_valid_z}), we can assume that the IV values are effectively randomized. As a result, the values of the treatment assignment generated by IVs, denoted $D_{IV}$, are also randomized. It follows that the estimate using $D_{IV}$, $\hat{\beta}_{IV}$, is theoretically free of unobserved confounding.

Complexity arises in using IVs mainly because $D$ and $D_{IV}$ are not technically the same variable. This means that ${\beta}_{IV}$ is a different estimand than $\beta$ and the IVs cannot be used to directly calculate the ATE but, rather, the "Local Average Treatment Effect" (LATE) where we are "local" to variation in the IVs.\citep{imbens_identification_1994} Fortunately, if there is no treatment effect heterogeneity and the assumptions for an IV are met, $\hat{\beta}_{IV}$ is consistent for $\beta$. 

As a simple demonstration of the above notions, the form of the LATE with a binary IV and binary treatment can be given by the Wald estimand in Eq. \ref{eqn:wald_equation}:

\begin{equation}
    \label{eqn:wald_equation}
    \beta_{IV} = \frac{E[Y|Z = 1] - E[Y|Z = 0]}{E[D|Z = 1] - E[D|Z = 0]}.
\end{equation}

Besides adding more intuition behind IV-derived treatment effects, this equation introduces the importance of IV strength or "predictive power" captured by $E[D|Z = 1] - E[D|Z = 0]$ or $\alpha$ in Figure \ref{fig:dag_simple}. Heuristically, when $\alpha$ is small then the IV is "weak" and if $\alpha$ is sufficiently large then the IV is "strong." In the linear setting, the finite sample bias of $\hat{\beta}_{IV}$ is partially a function of $\alpha$ where we incur large bias for $\beta$ with small values of $\alpha$.\citep{bound_problems_1995} For instance, in Eq. \ref{eqn:wald_equation}, the fraction is inflated.

The impacts of weak IVs are not just limited to finite samples. Recall that we cannot confirm we have a true IV using observed data and so we must assume our IV estimate is inconsistent for $\beta$. In this case, as an IV becomes weaker, the sensitivity of the corresponding estimate $\hat{\beta}_{IV}$ to IV independence assumption violations increases.\citep{wooldridge_econometric_2010} To elucidate this, suppose we had two IV candidates $Z_1$ and $Z_2$ with corresponding strengths $\alpha_1$ and $\alpha_2$, where $|\alpha_1| > |\alpha_2|$. For the same degree of violation in the independence assumptions (e.g. in Figure \ref{fig:dag_z_violation} $\delta = c > 0$ where $c$ is some constant) the inconsistency of an estimate derived from $Z_2$ would be greater than from using $Z_1$.

\begin{center}
\captionof{table}{Potential Treatment Assignment} \label{tab:potential_trt} 
\begin{tabular}{ p{1.5in} p{1in} p{1in}}\toprule[1pt]
\bf Sub-population & $D(0)$ & \bf $D(1)$ \\\midrule
Always Takers&  1 & 1\\
Compliers & 0 & 1 \\
Defiers & 1 &  0\\
Never Takers &  0 &  0\\
\bottomrule[1.25pt]
\end {tabular}\par
\end{center}

When there is treatment effect heterogeneity, even when we have a valid IV, $\hat{\beta}_{IV}$ can be inconsistent for $\beta$ because $\beta_{IV}$ is an estimand for a subset of the original population. Table \ref{tab:potential_trt} summarizes four distinct sub-populations related the IVs: always-takers, compliers, defiers, and never-takers. We can use potential outcomes once again but for treatment assignment: let $Z$ be a binary instrument and $D(0)$ be the treatment assignment had the value of the IV been 0 and $D(1)$ had the values of the IV been 1. 

In Table \ref{tab:potential_trt}, it is clear that changing values of the IVs results in changing values of the treatment assignment only for compliers and defiers. Therefore, we cannot identify always-takers and never-takers using IVs. In addition, we must impose a further assumption that the defier population does not exist for a given IV. This notion is called "monotonicity" where $D(1) \ge D(0)$ or vice-versa. Thus, we conclude that the subpopulation identified by the IVs are the compliers. To explain why we require monotonicity, we can rewrite Eq. \ref{eqn:wald_equation} as\citep{angrist_mostly_2009}

\begin{align*}
    E[Y|Z = 1] - E[Y|Z = 0] &= E[\big(Y(1)-Y(0)\big)\big(D(1) - D(0)\big)]\\
    &= \bigg(E[Y(1)-Y(0)|D(1) > D(0)]P(D(1) > D(0))\\
    &- E[Y(1)-Y(0)|D(1) < D(0)]P(D(1) < D(0))\bigg).
\end{align*}

Because of treatment effect heterogeneity, the ATE is differential depending on the subpopulation and there is the potential for a non-zero treatment effect in each group to cancel out. Of course, this does not occur if $P(D(1) < D(0)) = 0$ (no defiers) or $E[Y(1)-Y(0)|D(1) > D(0)] = E[Y(1)-Y(0)|D(1) < D(0)]$ (no treatment effect heterogeneity). 

Through a similar derivation in the denominator of Eq. \ref{eqn:wald_equation} as above, we arrive at a clearer definition of the LATE: $\beta_{IV} = E[Y(1)-Y(0)|D(1) > D(0)]$ or the ATE for compliers.\citep{imbens_identification_1994} This LATE, changes with chosen IV. If there are multiple IVs, then the LATE is a weighted average of LATEs characterized by each IV. When we have covariates included to establish the validity of $Z$ or decrease error in predicting $D$, then the LATE is an estimand defined on a population conditional on these covariates. Furthermore, unless the model is saturated, always and never-takers are included.\citep{abadie_semiparametric_2003,blandhol_when_2022} As most models in practice include covariates, the interpretability of IV models can be nebulous. 

In practice, there usually exists treatment effect heterogeneity so before using IVs we must determine if it is reasonable to target the LATE as a proxy for the ATE estimand. Consider the following scenarios. First, when treatment effect heterogeneity is unrelated to the choice of treatment then the estimates for compliers will not be systematically different from the other subpopulations. Next, when the first-stage is strong, the population characterized by the IVs will be relatively close to the overall population of the study. For example, if we have found a strong genetic determinant of some condition that was a valid IV (i.e. a Mendelian randomization strategy), it is plausible that, for the vast majority of the population, the occurrence of the condition would vary with the assignment of the gene. It follows that if the IV is weak, the LATE will only capture a small subset of the original population, introducing significant inconsistency in estimating the ATE. Lastly, there is a developing set of literature that relaxes the assumptions for the Wald estimand to equal the ATE such as requiring heterogeneity in the outcome caused by the treatment assignment to be independent from heterogeneity in treatment assignment caused by the IV as well as the IV itself.\citep{aronow_beyond_2017,wang_bounded_2018,hartwig_average_2020}

Identifying potential IVs is significantly less straightforward than identifying confounders, which is a main limitation of the approach. While there is usually abundant literature on predictors of a medical condition, the factors that determine the assignment of a treatment are difficult to study and are not usually studied. One reason for writing this paper is to generate expose  biomedical researchers to IVs such that potential IVs can be shared in the literature similar to how predictors are. In a similar vein to the confounder adjustment approach, we may begin by determining factors that predict treatment assignment and then prune those that affect the outcome. Determining confounders first is helpful as variables that were once invalid IVs may become valid after holding certain confounders constant. 

One popular source of IVs is variation in medical practice as it is well known that practice differs across physicians and regions across a wide variety of medical conditions.\citep{wennberg_dealing_1984,wennberg_small_1973,corallo_systematic_2014} If appropriate, we could use factors such as regional variation, facility prescribing patterns, attitudes to certain contraindications, physician preference, and calendar time as IVs. \citep{chen_use_2011,brookhart_preference-based_2007} For example, with access to the relevant data, physician preference can be quantified by tabulating the proportion of patients under each physician who were prescribed the treatment of interest. Following this, we can use these proportions to predict which treatment a new patient who sees any of these physicians will receive.

Even still, the validity of prescriber preference as an IV can be questioned. It could be that certain types of patients tend to select a physician that they know is more likely to give them the treatment (graphically, Figure \ref{fig:dag_z_violation} edges from $Z$ to $U$). Furthermore, geographic variation in general population health could necessitate higher utilization of treatments in some regions compared to others. Herein lies the value of identifying confounders in IV analyses: perhaps controlling for patient characteristics will block these pathways and greatly reduce assumption violations (e.g. Figure \ref{fig:cond_valid_z}). One takeaway, however, is that IV analysis can easily suffer from issues related to unobserved confounding.

Given a set of IVs, we should characterize each subpopulation. For medical practice patterns, most likely, some patients would not comply with a doctor's opinions; some patients could insist to get the treatment (alway-takers) and others would refuse under all circumstances (never-takers). One that does that opposite of what the doctor says (defiers) is possible and we will have to assume that they do not exist, which is practically untestable but can be reasoned as unlikely. Under this assumption, the LATE would roughly be those who follow the doctors' orders. All of this considered, the analyst should determine whether the complier treatment effect is of scientific value.

\subsection{Interactions of the Confounder and IV Approaches}

The confounder and IVs approaches are deeply related. Therefore, even if an analyst decided to pursue one approach over another, awareness of the principles of the other approach is important. One pervasive issue in this vein is adjusting for an IV as if it was a confounder. Widely-cited guidelines such as Hirano and Imbens (2001) state that variables that are predictive of treatment assignment should be selected for confounder methods like propensity scores,\citep{hirano_estimation_2001} which risks adjusting for IVs and mediators. In the best case, treating IVs as confounders decreases precision because it does not explain variation in the response. Even worse, when there is unobserved confounding, existing inconsistency is amplified.\citep{bhattacharya_instrumental_2007,wooldridge_should_2016,ding_instrumental_2017,pearl_class_2010} By adjusting for IVs, we reduce variation in the treatment that is uncorrelated with the unobserved confounding. Thus, variation in the treatment produced by unobserved confounding proportionally increases, which causes more bias in the treatment effect. 

The impact of adjusting for IVs and mediators demonstrates why one should avoid a purely "kitchen sink," data-driven approach to variable selection for causal inference. Simply because the estimate of the treatment effect changes when a variable is introduced does not necessarily mean it should be adjusted for. This is one reason why we advocate that confounders largely be sourced \textit{a priori} by first hypothesizing predictors of the outcome. If one is reasonably certain that a variable is predictive of the outcome but is unlikely to be associated with the predictor of interest, one has a "precision variable," which still may be of use. Specifically in the linear model setting, adjusting for such a variable will decrease standard errors in the treatment effect estimate with no cost to bias.\citep{brookhart_variable_2006,greenland_confounding_1999} 

\section{Methodology for Estimating Causal Effects}

In this section, we discuss three popular approaches for estimating causal effects: regression adjustment using ordinary least squares (OLS) or generalized linear models, propensity score weighting with inverse probability of treatment weighting (IPTW), and utilizing two-stage least squares (2SLS) with IVs. Regression and IPTW are used in the confounder approach while 2SLS serves as a methodology for the IV approach. Although there are many other methods such as g-computation, targeted learning, proximal causal
learning, non-parametric 2SLS, and two-stage residual inclusion, we will spend most of our discussion
on regression, IPTW, and 2SLS because they are most commonly used and well-studied. We briefly speak about recent advances in later sections.

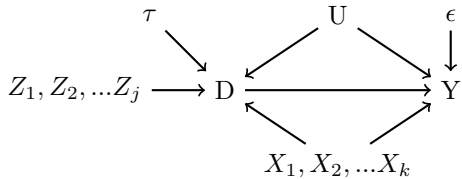
\begin{figure}[h]
    \centering
    \begin{tikzpicture}[thick]
        \node (1) at (0,0) {$Z_1,Z_2,...Z_j$};
        \node (2) at (2,0) {D};
        \node (3) at (5,0) {Y};
        \node (4) at (3.5,-1) {$X_1,X_2,...X_k$};
        \node (5) at (3.5,1) {U};
        \node (6) at (1,1) {$\tau$};
        \node (7) at (5,1) {$\epsilon$};
        
        \path [->] (1) edge (2);
        \path [->] (2) edge (3);
        \path [->] (4) edge (2);
        \path [->] (4) edge (3);
        \path [->] (5) edge (2);
        \path [->] (5) edge (3);
        \path [->] (6) edge (2);
        \path [->] (7) edge (3);
    \end{tikzpicture}
    \caption{A DAG with more IVs ($Z$'s) and observed confounders ($X$'s) as well as relevant stochastic errors $\tau$ and $\epsilon$ for $D$ and $Y$, respectfully}
    \label{fig:larger_DAG}
\end{figure}

A more sophisticated version of Figure 1 is presented in Figure 5. We have added a vector of IVs of length $j$ ($Z_1,Z_2,...Z_j$) and a vector of observed confounders of length $k$ ($X_1,X_2,...X_k$). In addition, we have there are now stochastic errors for $D$ and $Y$. For simplicity, assume the effect of each IV is the same magnitude and similarly for each confounder and that the outgoing arrows capture from the joint effect. Furthermore, $U$ captures all unobserved confounding though, in reality, there are likely many variables. Assuming the relationships between variables are linear, we can write the following system of relevant structural equations:

\begin{equation}\label{eqn:first_stage}
        d_i = \gamma_0 + x_i^{T}\gamma_X + z_i^T\gamma_Z + u_i\gamma_U + \tau_i,
\end{equation}
\begin{equation}\label{eqn:second_stage}
        y_{i} = \beta_0 + \beta_D d_i + x_i^{T}\beta_X + u_i\beta_U + \epsilon_i.
\end{equation}

Eq. \ref{eqn:first_stage} depicts the treatment assignment or "first stage" while Eq. \ref{eqn:second_stage} depicts the outcome or "second stage." The estimand of interest is $\beta_D$. Because the treatments are usually binary variables, the functional form of the treatment assignment is commonly characterized using a logit or probit model. For ease of exposition, however, we will assume a linear probability model (LPM) as in Eq. \ref{eqn:second_stage}. Please see later sections for discussion on the use of LPMs for modeling binary treatment assignment.

Using the above two equations, we can provide a high-level overview of the three methods in practice. Regression methods fit Eq. \ref{eqn:second_stage} with the treatment and the observed confounders to estimate $\beta_D$. $U$ is not observed so the estimate is inconsistent due to misspecification. Meanwhile, IPTW first fits Eq. \ref{eqn:first_stage} with only the confounders to predict the propensity score for all subjects. The propensity scores will be used to compute a weighted sum that will allow us to estimate $\beta_D$. Because $U$ is missing, the predicted propensity scores will not be correct nor adequate to achieve ignorability conditional on the propensity score.

2SLS will fit Eq. \ref{eqn:first_stage} as stated (except for $U$) and use the predictions to construct $\hat{D}$. We then effectively substitute $D$ in Eq. \ref{eqn:second_stage} and fit an OLS model to estimate the coefficient in front of $\hat{D}$. Importantly, the omission of $U$ does not affect the consistency of this estimate under the conditions of Figure \ref{fig:larger_DAG} and if there is no treatment effect heterogeneity then we have a consistent estimate of $\beta_D$. If there is then, at the least, the estimate is not affected by $U$.

\subsection{Regression Adjustment and Propensity Score Methods}

While OLS and IPTW are different mathematically, they are conceptually similar: both seek to isolate variation in the outcome caused by the treatment by eliminating variation caused by confounding factors. Regression adjustment can be thought of as blocking paths in Figure \ref{fig:larger_DAG}, which is another way of stating that we are holding $X$ constant in order to isolate the effect of $D$ on $Y$ and obtain the direct causal pathway with the following estimand:
\begin{align*}
    \beta_D^{OLS} = E[Y|D=1,X=x] - E[Y|D=0,X=x].
\end{align*}

On the other hand, IPTW weights outcomes based on the probability of receiving ($p(X) = P(D=1|X)$, creating a pseudo-population that balances confounders across the treatment groups in a similar rationale to randomization. IPTW has the following estimand:

\begin{equation}
    \label{eqn:iptw}
    \beta_D^{IPTW} = E\bigg[\frac{DY}{p(X)}\bigg] - E\bigg[\frac{(1-D)Y}{1-p(X)}\bigg].
\end{equation}

For a more concrete example of how a pseudo-population is constructed, suppose an individual in the treatment group had a propensity score of 0.1. In other words, this individual is very likely to receive the control and has many similar subjects in the control group. The weighted outcome of this individual can represent the counterfactual outcomes of the comparable control group subjects and so we would weight that individual's contribution to the treatment effect ten times. As a consequence of such a procedure, in Figure \ref{fig:larger_DAG} the pseudo-population will theoretically no longer contain edges for the $X$'s because the treatment assignment cannot be explained by covariate imbalance. A balance of observed confounders, however, does not imply a balance of unobserved confounders: the act of balancing observed confounders may increase the imbalance of these unobserved confounders.\citep{brooks_squeezing_2013}

Propensity scores are often used to match individuals across treatment groups. Once one or more suitable matches are found for each subject in the treatment groups, we can compute the differences in outcomes, and average them to obtain the average treatment effect on the treated (ATT), $E[Y(1) - Y(0)|D = 1]$. We focus on IPTW and not propensity score matching because, under unobserved confounding, the ATT will not equal the ATE:

\begin{align*}
    \beta^{ATT} &= E[Y(1) - Y(0)|D = 1]  = E[Y(1)|D = 1] -  E[Y(0)|D = 1]\\
    &\ne  E[Y(1)|D = 1] -  E[Y(0)|D = 0]  \text{  ($Y(0) \notindependent D$)}\\
    &= \beta^{ATE}.
\end{align*}

While our discussion of methodology in this paper mainly centers around the ignorability assumption, it is important to briefly touch upon the implications of positivity assumption violations. In propensity score methods this means each subject has a positive probability of receiving the treatment given each level of the covariates or $0 < P(D=d|\boldsymbol{X} = \boldsymbol{x}) < 1$ for all $d \in D$ and $\boldsymbol{x} \in \boldsymbol{X}$. Another way to conceptualize the positivity assumption is the notion of "common support" where there must be full overlap in each group's distribution of propensity scores or, by extension, their observed covariates. Even when there is no unobserved confounding, positivity violations can arise when we fail to observe certain variables that are needed to create overlap. Therefore, for the subpopulations lacking overlap, extrapolating counterfactual claims can lead to erroneous conclusions. 

Conceptually, a violation of the positivity assumption means that we are dividing by 0 in Eq. \ref{eqn:iptw}. In practice, this results in extreme weights, which both increases variability in the parameter estimates but also impacts finite sample bias because the estimate is weighted towards the few extreme observations.\citep{petersen_diagnosing_2012}. In addition, near-positivity violations, or individuals who are extremely unlikely to receive the treatment or placebo, pose similar issues for estimation. When using OLS, positivity violations pose a similar bias of the estimand as IPTW. Nevertheless, OLS does not face the same degree of finite sample estimation issues as IPTW when there are positivity or near-positivity violations. 

While IPTW and OLS target the same estimand, the ATE, and both are vulnerable to inconsistency via unobserved confounding, there are differences to consider in practice. If there are no extreme weights, by encapsulating many covariates in the propensity score, IPTW will generally be more efficient than OLS because of the degrees of freedom saved. If there are extreme weights, however, the instability of the variance of estimators will be larger than that of adjustment-based regression methods.

One common solution to extreme weights in propensity score methods is to trim extreme weights. This procedure, however, risks estimating the treatment effect for a population different than the original target population. In other words, for a decrease in variance, there is a potential increase in bias. Furthermore, the direction of this bias is difficult to determine because one must define a new population resulting from truncation. Though one could argue that the bias due to positivity violations could be advantageously traded-off with the bias due to truncation\citep{westreich_invited_2010}. Another common solution to extreme weights could be to use stabilized weights as opposed to conventional inverse probability weights.\citep{robins_marginal_2000}

OLS and IPTW also have potential differences in ease of reproducibility and interpretability. Because propensity scores are the result of fitting a model for treatment assignment in order to generate propensity scores, methods ranging from logistic regression to random forests may be used. An issue, however, arises when different models produce different sets of propensity scores resulting in different pseudo-populations. This poses challenges for reproducibility across different studies of the same population. The simplicity of OLS arguably reduces the risk of this since basic adjustment can be easily communicated. On the other hand, when unaccounted for treatment effect heterogeneity exists, OLS will generate a marginal treatment effect estimate that is implicitly weighted by the covariance structure in the observed data sample as opposed to explicit weighting in IPTW.\citep{aronow_does_2016,angrist_mostly_2009}

One advantage of the propensity score is that data-driven selection of the confounders to model treatment assignment is done separately from the fitting of Eq. \ref{eqn:second_stage}. In contrast, adding and removing confounders in OLS in a data-driven fashion will also affect the estimand, estimate, and corresponding inference for the treatment effect. Therefore, IPTW is able to control inflation in Type I error from repeated testing of the treatment effect coefficient as a result of fitting several models.

Though it may be tempting to cast estimating propensity scores as a prediction problem, this may lead to unintended consequences. The original philosophy of propensity scores from Rosenbaum and Rubin is not to fit the first-stage as well as possible; rather, it is to find a balancing score sufficient to achieve ignorability.\citep{rosenbaum_central_1983,austin_moving_2015,vegetabile_optimally_2020,imai_covariate_2014} Furthermore, measures of model performance like the C-statistic do not provide useful information to suggest unobserved confounding is mitigated more in one model than another.\citep{weitzen_weaknesses_2005}. IVs are predictors of treatment assignment and, yet, they are adverse to causal estimation if included in the model.\citep{bhattacharya_instrumental_2007,austin_moving_2015} In addition, including variables that are predictive of the outcome but not the treatment can help improve efficiency of treatment effect point estimates.\citep{brookhart_variable_2006}. Therefore, we suggest that variable selection for propensity score modeling is not conducted purely by optimizing out-of-sample model prediction error.

In contrast, the simplicity of using one equation in regression methodology assists in avoiding confusion between prediction and inference goals because there is no intermediate step in obtaining an estimate for $\beta_D$. Indeed, an optimized model MSE may reduce coefficient standard errors and yet could lead to issues in internal validity. For example, if we fit Eq. \ref{eqn:second_stage} with LASSO to select confounders that optimized out-of-sample prediction error, the elimination of confounders via shrinkage to zero could potentially induce further omitted variable
bias due to no longer conditioning on an observed confounder.\citep{chernozhukov_post-selection_2015}

A combination of the IPTW and regression methodology is the augmented IPTW (AIPTW) with the so-called "doubly robust" property: a consistent estimate is obtained if either the propensity score (Eq. \ref{eqn:first_stage}) or the outcome equation (Eq. \ref{eqn:second_stage}) are correctly specified. Certainly, AIPTW offers more robustness than IPTW but its practical advantage is unclear. Firstly, in the case where one suspects the propensity score equation is misspecified but the outcome equation is not, OLS also will result in a consistent estimate and could theoretically be more efficient. Secondly, an unobserved confounder would cause misspecification in both equations, rendering any estimate inconsistent. As such, it is unclear how the inconsistency due to unobserved confounding in the AIPTW compares to that of IPTW or OLS. 

\subsection{Instrumental Variable Methods}

2SLS is the most commonly used and well-studied IV method. The motivations of 2SLS largely stem from the inadequacy of OLS to provide a consistent estimate of $\beta_D$. Under the reduced form in Eq. \ref{second_stage_reduced}, omitting $U$ leads to inconsistency because $\alpha_D = \beta_D + \frac{Cov(D,U)}{Var(D)}$.
\begin{equation}\label{second_stage_reduced}
    (y_{i,t=2} - y_{i,t=0}) = \alpha_0 + \alpha_D d_i + x_i^{T}\alpha_X + \phi_i
\end{equation}

Another name for this scenario is that there is "endogeneity" or correlation of $D$ with the error term. This is because $\phi_i = \beta_U + \epsilon_i$, which is correlated with $D$ and $E[\phi_i|D,X] \ne 0$. IV methods will identify a re-characterized treatment, $D_{IV}$, using exogenous variation (uncorrelated with the error term) such that $Cov(D_{IV},\phi_i) = 0$. 

In the first stage of 2SLS, we regress the $X$'s and $Z$'s (design matrix $\boldsymbol{F}$) on $D$ and use the projection matrix $P_Z = \boldsymbol{F}(\boldsymbol{F}^T\boldsymbol{F})^{-1}\boldsymbol{F}$ to obtain predicted values $\hat{D}_{IV}$. In the second stage, we use the $P_Z$ to obtain the coefficients $\boldsymbol{\beta} = (\boldsymbol{S}^TP_Z\boldsymbol{S})^{-1}\boldsymbol{S}^TP_ZY$ with design matrix $\boldsymbol{S}$ consisting of the $X$'s and $D$. By including the $X$'s in both the first and second stages, we can improve the prediction of $D$ and covariates serve as their own IVs. If all assumptions are met, and there is no treatment effect heterogeneity, then $\beta_D^{IV}$ is consistent for $\beta_D$ but not necessarily unbiased. In fact, in the case where we only have one IV for one endogenous variable, the first moment does not exist.\citep{kinal_existence_1980} 

There are considerable trade-offs in the IV analysis: consistency comes at the cost of increased standard errors compared to OLS since we only use the exogenous variation in the treatment and, thus, have less "information" to calculate the treatment effect.\citep{wooldridge_econometric_2010} In the case where the IVs are weak, the finite bias will move towards OLS as weakness increases (i.e. first-stage coefficients go towards 0) and inflate estimator standard errors.\citep{angrist_split-sample_1995,bound_problems_1995} This is because our treatment effect is determined only by compliers, a subset of the overall population. If the IVs are correlated with the second stage error term (i.e. $U$) not only will the estimates be inconsistent but the magnitude of inconsistency is greatly affected by the IV strength.\citep{wooldridge_econometric_2010} This can be observed by deriving the form of the 2SLS estimand under this condition. 

\begin{equation}\label{eq:iv_consistency}
    \beta_{D}^{IV} = \beta_D + \frac{Cov(Z,\epsilon)}{Cov(Z,D)}.
\end{equation}

$Cov(Z,\epsilon)$ captures the degree of the violation in the independence of the IV, and $Cov(Z,D)$ captures the first stage strength. Rewriting covariances as correlations in Eq. \ref{eq:iv_consistency} and in the OLS estimate $\alpha_D$, we obtain:

\begin{align*}
    \beta_{D}^{IV} = \beta_D + \frac{\sigma_\phi}{\sigma_D}\frac{Corr(Z,\phi)}{Corr(Z,D))},
\end{align*}
\begin{align*}
    \alpha_D = \beta_D + \frac{\sigma_\phi}{\sigma_D}Corr(D,\phi).
\end{align*}

It is clear that when the IVs are weak, 2SLS inconsistency can be greater than even that of OLS if $|Corr(D,\phi)||Corr(D,Z)| < |Corr(Z,\phi)|$. Considering that we can never confirm the IV independence assumption, the use of weak IVs may be perilous.

Resuming our assumption of valid IVs, another perhaps helpful perspective is that the first-stage is chiefly a prediction task. In this interpretation, weak IVs lead to inaccurate first-stage predictions, which leads to finite sample bias because we are unable to adequately capture the treatment assignment of the original population of interest. Simply adding more weak IVs to the first-stage rarely improves the issue; indeed, packing the first-stage with too many instruments will lead to overfitting and, hence, finite sample bias.\citep{roodman_note_2009} These issues are partially mitigated by large sample sizes. 

Unlike many of the assumptions discussed, the degree of instrument relevance is somewhat determinable using the data. Because 2SLS utilizes OLS for the first stage, the F-statistic is commonly used to measure the joint strength of the IVs with 10 being a "rule of thumb" for sufficient strength. In the case of heteroskedasticity, a robust F-statistic can be used but variance estimates may be noisy.\citep{young_consistency_2022} Nevertheless, using a sample statistic to infer upon assumptions could be problematic. For instance, Young 2017 points out that there is a relatively high chance of spuriously obtaining a high F-statistic and unreliability of "guaranteed bounds" of size and bias of weak IVs tests. In addition, weak IV tests assume that the IVs are valid in the first place or else coverage will be incorrect.\citep{guggenberger_asymptotic_2012}

Data-driven fitting of the first-stage may prompt concerns about external validity. In theory, we could select variables in the data that optimized a cross-validated F-statistic. In this process, however, we would fail to address the impacts on the interpretability of the LATE. This scenario introduces difficult situations where one set of IVs could have a worse F-statistic but is more interpretable for the desired use case. This is why we advocate for first thinking through the conceptual soundness of the selection of IVs given a set of theoretically strong predictors of the treatment.

Although the independence of the IVs is untestable, there are a few interesting falsification tests that one could employ. One straightforward test is to compare the values of potential confounders across values of the IVs similar to how one examines values of potential confounders across levels of the treatment.\citep{baiocchi_tutorial_2014} Imbalance of IVs across an observed confounder is problematic because the observed confounders could be related to an unobserved confounder. Another test that assumes observed confounders may be related to unobserved confounders involves negative control outcomes, or populations constructed to falsify IV independence.\citep{davies_how_2017} Of course, these tests cannot ensure IV assumptions are met and are subject to data availability.

In the multiple IV case, given there is one valid IV, one may test if additional IVs influence the outcome through Sargan's \textit{J}-statistic. \citep{Sargan_estimation_1958} By running OLS on the residuals of 2SLS on the IVs, if one or more coefficients are not 0, we have some sort of violation. This test may have deceivingly high p-values and poor power when one IV is weak but valid and the others are strong but invalid.\citep{kiviet_instrument_2021} The scenario of a mix of strength and validity of IVs poses an interesting question about whether one would choose a strong but slightly invalid IV over a weak but valid IV.

\begin{center}
\captionof{table}{Comparison of Regression, IPTW, and 2SLS under Unobserved Confounding} \label{tab:comp_reg_iptw_2sls} 
\begin{tabular}{p{0.75in} p{1.5in} p{1.75in} p{1.75in}}\toprule[1pt]
& \bf Regression & \bf IPTW & \bf 2SLS\\\midrule
\bf Mechanism &
Hold constant variables $\leftrightarrow$ Block paths on DAG& 

Balance groups via over-weighting "rare" individuals $\leftrightarrow$ Eliminate arrows in DAG via independence&
Identify treatment from variation independent from unobserved confounding $\leftrightarrow$ 
Create a new DAG with no backdoor paths for treatment

\\\midrule
\bf Consistency \\\bf Assumptions &
\begin{itemize}
    \item SUTVA
    \item No unmeasured confounders
    \item Positivity
\end{itemize}& 
\begin{itemize}
    \item SUTVA
    \item No unmeasured confounders
    \item Positivity and No Near-Violations
\end{itemize}&
\begin{itemize}
    \item SUTVA
    \item IVs are predictive of treatment assignment
    \item IVs are as good as randomized
    \item Does not directly affect outcome
    \item (For ATE) no treatment effect heterogeneity
    \item (For LATE) monotonicity 
\end{itemize}
\\\midrule
\bf Considerations &
\begin{itemize}
    \item Inconsistency amplification with inclusion of IVs
    \item Treatment effect heterogeneity: conditional effect $\ne$ marginal effect
\end{itemize}& 
\begin{itemize}
    \item Inconsistency amplification with inclusion of IVs
    \item Fitting propensity score model mistaken for prediction task
    \item With higher dimensionality, difficult to interpret positivity violations
    \item Balancing on observed confounders does not imply balance on unobserved confounders
    \item Treatment effect heterogeneity: estimates marginal effect
\end{itemize}&
\begin{itemize}
    \item Inefficiency compared to confounder methods
    \item Validity possibly contingent on unobserved confounders
    \item Treatment effect heterogeneity: LATE $\ne$ ATE
    \item LATE difficult to interpret for many IVs, weak IVs, and inclusion of covariates
    \item Weak IVs: increased finite bias and increased sensitivity to validity violations
\end{itemize}\\
\bottomrule[1.25pt]
\end {tabular}\par
\end{center}

\section{Complexities that Arise with Non-Linearity}

So far, our discussion has been focused on the case where the outcome is continuous and, thus, we can reasonably assume linear structural equations. When the outcome is not continuous, some of the statements we have previously made must be modified. In particular, we will revisit the consequences of adjusting for confounders, IVs, and precision variables, marginal versus conditional estimands, and the use of 2SLS for binary treatments and non-continuous outcomes.

First, in the non-linear model setting the estimand corresponding to the treatment effect will change by including not only confounders but also precision variables because of non-collapsibility.\citep{greenland_confounding_1999} Mathematically, because the covariates are encapsulated in a non-linear function (e.g. a link function), after adjusting for a precision variable, we cannot simply distribute the expected value such that we recover the "before adjustment" treatment effect.\citep{pearl_causality_2009} A further consequence is that adjusting for any variable will increase coefficient standard errors of the variables already present. For instance, in logistic regression, the unexplained variance must stay fixed so the explained variance will increase upon adjustment for new variables, leading to coefficient values increasing in magnitude.\citep{mood_logistic_2010,schuster_noncollapsibility_2021} Adjusting for precision variables will increase standard errors of the treatment effect but slightly increase the power to reject the null of no effect; this is due to the magnitude of the adjusted point estimate increasing relative to slight increases in the standard error of the estimate.\citep{schisterman_overadjustment_2009}

The presence of non-collapsibility means that the interpretation of coefficients before and after adjusting for variables differs even in cases where there is independence between the adjustment variable and the treatment variable. Marginal estimates, without confounders, will therefore be different from conditional (on the confounders and precision variables) estimates. Comparing methodologies, IPTW will give a marginal estimate whereas regression will give a conditional estimate.\citep{schuster_noncollapsibility_2021} For example, after adjusting for the confounders, IPTW produces the odds ratio for the population characterized by the sample while a logistic regression model would compute the odds ratio for someone with the average value of the confounders.\citep{karlson_marginal_2021} In the linear case, the difference between conditional and marginal effects was only present under treatment effect heterogeneity.

Whether the marginal or conditional estimand is preferable depends on the scenario. One could argue that the conditional estimand is more applicable to settings where a physician is already conditioning upon knowledge of various patient characteristics like sex, age, and comorbidities. In addition, conditional estimands may be better transported to other populations such as future populations. On the other hand, marginal estimands can be more interpretable and comparable across studies.\citep{vansteelandt_invited_2011} They may also be preferred when the covariates potentially included in the model are not easily observed or measured in practice.

The issue with using IVs as confounders persists in the non-linear setting: adjusting for IVs as if they were confounders amplifies existing bias none and could even introduce bias where none previously existed.\citep{pearl_class_2010,ding_instrumental_2017} Amplification is for the same reasons as the linear setting while introducing bias occurs when the IVs are dependent on the outcome given the treatment.\citep{pearl_confounding_2014} Though Ding et al. do find specific situations where this bias was not present under their proposed monotonicity conditions for the treatment selection and outcome model.

A primary challenge for standard IVs methods is that 2SLS misspecifies a non-linear functional form in the case where we have a binary treatment. Nevertheless, analysts still may utilize 2SLS in non-linear settings such as through LPMs. Simulations have shown that LPM can produce low inconsistency in the estimates of the LATE.\citep{basu_2sls_2018} A counterpart of 2SLS, two-stage residual inclusion (2SRI), where one takes the residuals from the first stage as a covariate in the second stage, did not perform nearly as well. Furthermore, claims that non-linear 2SRI, (using a probit model for example) are able to recover the ATE (as opposed to the LATE) are questionable.\citep{chapman_treatment_2016} Quantifying the effect of IV method misspeficiation relative to the confounder methods is an avenue of research.

Intuitively, using LPMs appears inappropriate as there is nothing preventing prediction of values that are outside of the interval $[0,1]$, which leads to bias and inconsistency of LPM estimates. However, if the population of interest only has probabilities between a certain range not close to 0 or 1, then this will seldom be an issue because we will not have errant predicted values. Thus, LPMs will be consistent and unbiased.\citep{horrace_results_2006} One could also argue if there are true probabilities that are 0 or 1, then we have a positivity violation and propensity score methods, whether they use a logit, probit, or LPM to calculate propensity scores, will also run into issues. One solution for any method is to truncate probabilities but this will be at the cost of bias for the original causal effect. 

We can rarely mimic the 2SLS substitution procedure with non-linear models, such as two-stage probit, and maintain consistency for the treatment effect (such an action is called "forbidden regression").\citep{wooldridge_econometric_2010} Instead, one can consider a three-step procedure: fit the first-stage with a non-linear model, regress the predicted values on the treatment in OLS excluding the IVs, and, lastly, fit the second stage with a linear or non-linear model.

\section{Flexible Confounder and IV Methodology}

Recent developments in causal inference methodology have been mostly focused on relaxing assumptions on common approaches like OLS, IPTW, and 2SLS. These developments are predominately motivated by the wish to reduce the impacts of model misspecification and the view that many subtasks of estimation are primarily prediction-based, allowing for more flexible modeling using ML. Another motivation that we will not discuss is sparsity (e.g. incorporating more confounders or IVs than observations). We will cover three approaches: targeted minimum loss-based estimation (TMLE), Post-LASSO (PL), and double machine learning (DML).\citep{van_der_laan_targeted_2010,belloni_sparse_2012,chernozhukov_doubledebiased_2018}

TMLE improves upon AIPTW by modeling nuisance quantities flexibly in addition to achieving double robustness. Nuisance quantities are needed to calculate the treatment effect but are not of direct interest to the research question. For example, the propensity score can be considered a nuisance quantity. By using ML and a "targeting step," we can avoid nuisance quantity misspecification and optimize the bias-variance tradeoff for the ATE.\citep{luquefernandez_targeted_2018,diaz_machine_2020} To estimate nuisance quantities, TMLE includes cross-validation-optimized weighting ensemble of ML algorithms (e.g. penalized regression, random forest, gradient boosted trees) and "cross-fitting." Cross-fitting bears some similarities to cross-validation in that "out-of-sample" data, or partitions of the data not used to fit models, is utilized to reduce overfitting. In addition, cross-fitting allows one to avoid proving complicated Donsker conditions for asymptotic normality.\citep{van_der_laan_unified_2003} In addition, TMLE offers some robustness in near-violations of the positivity assumption.

Briefly, the estimation procedure of TMLE is as follows. Let $D$ represent the assignment of a binary treatment and $\textbf{X}$ a set of confounders. First, perhaps using ML, we use the confounders to fit initial models for treatment assignment and outcome: $g(D,\textbf{X})$ and $\Bar{Q}^0(D,\textbf{X})$, respectively. Following this, we combine predictions from these equations along with a fluctuation parameter $\epsilon$ to give us double-robustness and local efficiency. Now, we will update the outcome equation to $\Bar{Q}^1(D,\boldsymbol{X})$ and then estimate the ATE using a "plug-in" estimator of the form 
\begin{align*}
    \hat{\beta}^{TMLE} = \frac{1}{n} \sum_{i=1}^{n} {\Bar{Q}^1(d_i=1,\boldsymbol{x_i})-\Bar{Q}^1(d_i=0,\boldsymbol{x_i})}.
\end{align*}

While TMLE takes a plug-in estimator approach, DML, and its precursor PL, utilizes estimating equations allowing them to compute both confounder and IV estimates. These methods utilize orthogonalization and the Frisch-Waugh-Lovell theorem to first perform regressions on the confounders and IVs and then use their residuals in subsequent estimating equations for the treatment effect that are "locally insensitive" to misspecification. Similar to TMLE, we utilize ML algorithms to fit the residual-creating regressions. The post-LASSO name comes from the fact that we first run LASSO regression to select variables and then refit OLS on the selected variables, which mitigates issues in LASSO variable selection. DML extends this notion of estimating nuisance quantities to more ML algorithms and includes double robustness. Asymptotic considerations, namely Donsker conditions for normality, are handled through cross-fitting.

For the IV model, we use the moment condition $E[\Phi_i(\beta_0,\eta_0)]=0$ where $\beta_0$ is the coefficient for the treatment effect and $\eta_0$ are the coefficients for the nuisance quantities based on the confounders and IVs. The score function takes the form $\Phi_i(\beta,\eta) = (\Tilde{\rho}_i^y - \Tilde{\rho}_i^y\beta)\Tilde{v}_i$, which is simply the canonical form of a generalized method of moments using $\Tilde{v}_i$ as an IV. The $\Tilde{\rho}_i^y$, $\Tilde{\rho}_i^y\beta$, and $\Tilde{v}_i$ terms are are all results of orthogonalization. For example, $\Tilde{\rho}_i^y =y_i - x_i^T\theta$ where $\theta$ was estimated via PL or LASSO. DML extends the orthogonalization procedure to other ML algorithms by assuming partially linear models.

The "robustness" to misspecification, whether via unobserved confounders or regularization of relevant confounders, is through the orthogonality condition:
\begin{equation}\label{eq:ortho_cond}
    \frac{\partial}{\partial \eta} E[\Phi_i(\beta_0,\eta)]\big|_{\eta=\eta_0} = 0, \forall \eta.
\end{equation}

Eq. \ref{eq:ortho_cond} essentially states that if there is some error in estimating the nuisance quantities, the resulting estimate will still be consistent for $\beta_0$. The authors call this "Neyman orthogonality" as the concept of "locally" robust estimates dates back to work done by Jerzy Neyman in 1959.\citep{neyman_jerzy_optimal_1959} The robustness of local insensitivity remains to be rigorously investigated in real world scenarios.

The fitting of the first stage of an IV analysis using ML and cross-fitting could result in inference more robust to weak IVs. In 2SLS, the estimated coefficients and, by extension, error term in the second stage is correlated with that of the first, resulting in "finite sample" bias. Thus, if ML can generate better predictions than OLS, we can decrease the first-stage error. In addition, we can utilize sample-splitting to break the association between the two error terms, which reduces the impact of weak IVs.\cite{angrist_split-sample_1995} We first partition the sample, for example in half, and use the first half to fit the first-stage and compute the predicted values for the treatment with the second half. The predictions based on second half are then used in the second stage. Cross-fitting expands on the idea of sample-splitting by repeating the process on the "unused" partition (i.e. partition used in the second stage) and then pooling the results across each fold. Both DML and TMLE use the cross-fitting procedure while PL only uses sample-splitting.

Theoretically, the benefits of ML and causal inference methods are clear but the extra complexity carries some caveats. There is a great danger of overfitting and spurious predictions, especially when the sample size is too low. For example, in predicting binary outcomes, one simulation study found that 20 to 50 "events" per variable were necessary for logistic regression to achieve stable prediction performance.\citep{van_der_ploeg_modern_2014} For more sophisticated methods like neural networks and random forests, instability persisted above 200 events per variable. Translating this to causal inference, because of overfitting, the external validity of ML prediction is questionable, especially when the proportion of treatment assignment is near 0 or 1 as even larger samples will be needed. Furthermore, ML methods could diminish the reproducibility of causal effects across studies and can be computationally expensive without much gain over "traditional" methods.

The in-practice performance of ML and causal inference of these algorithms is still under question. In one simulation study, Angrist and Frandsen were able to find meaningful gains in selecting confounders but not IVs using ML.\citep{angrist_machine_2022} One interesting phenomenon they highlighted was "pretest bias:" when LASSO decides whether to retain a coefficient or not, there is an implicit test being performed against some threshold contingent on the lasso regularization parameter and the sample size. Consequentially, when the true first-stage coefficients are zero or near-zero, bias in the treatment effect is introduced.\citep{andrews_weak_2019} In the paper, sample-splitting or cross-fitting mitigated pretest bias with ML not offering notable improvement in parameter estimation. The apparent importance of out-of-sample procedures highlights further needs for large sample sizes in sophisticated techniques. 

Perhaps the work on ML and causal inference, though innovative, "buries the lede," so to speak. Far and away, unobserved confounding is the greatest threat to the validity of observational studies. Though nuisance quantities can be more flexibly modeled, the bias due to omitted variables likely dwarfs potential model-fitting issues that arise when using traditional models such as OLS. Sophisticated methods are more robust to violations in the assumptions than their traditional counterparts but this does not eschew properly selecting confounders and IVs – evaluating the paradigm is paramount. For example, there has been some literature on estimation with some "imperfect" IVs that relax some variable selection techniques but these methods presuppose that one has at least some valid IVs.\citep{windmeijer_use_2018}.

In terms of external validity, in non-linear settings, it is also unclear whether these methods estimate a marginal or conditional effect. For the IV setting, if the first-stage is characterized by ML, where there is a significant amount of regularization, then the definition of the LATE can be vague. More perilous is the possibility for confounders that ensure conditional independence of the IVs to be omitted from the first-stage due to regularization. In PL, one could simply choose to not shrink to zero certain variables but for DML, it is unclear how this can be prevented in more sophisticated algorithms like random forests.

\section{Comparing Approaches for Applied Data Analysis}

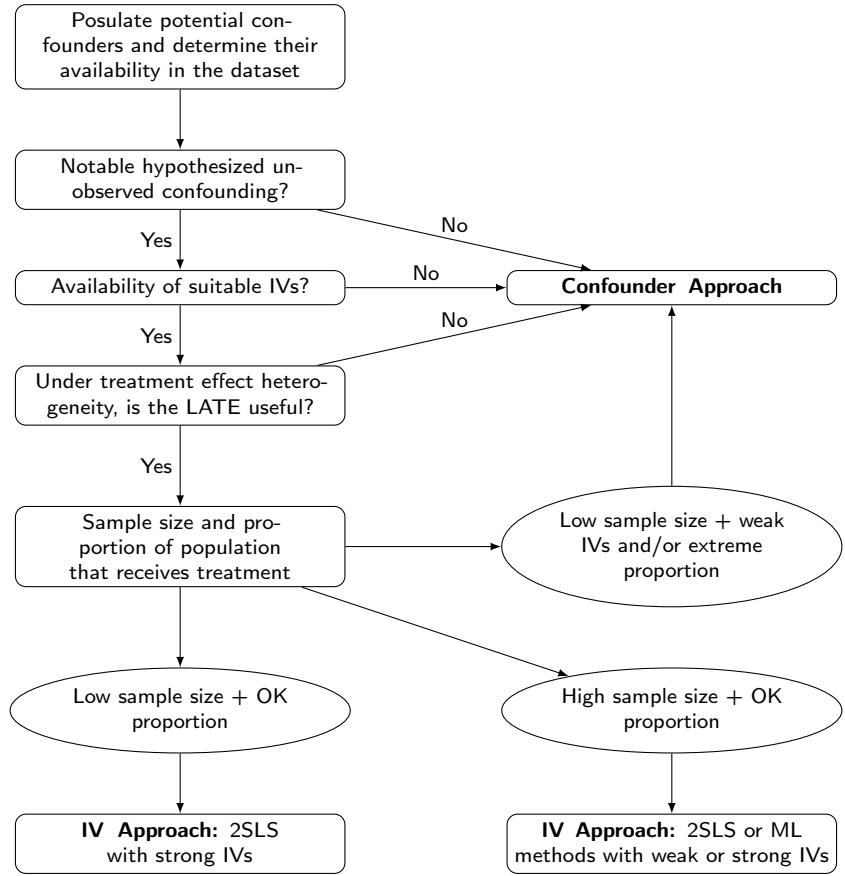
\begin{figure}[h!]
    \begin{center}
    \captionof{figure}{Flowchart presenting a possible heuristic for selecting an approach}
      \sffamily
      \footnotesize
      \label{fig:method selection}
    \begin{tikzpicture}
    \matrix (m)[matrix of nodes, column  sep=2cm,row  sep=8mm, align=left, nodes={rectangle,draw, anchor=center} ]{
        |[block]| {Posulate potential confounders and determine their availability in the dataset} & \\
        |[block]| {Notable hypothesized unobserved confounding?} &  \\
        |[block]| {Availability of suitable IVs?}          &   |[block]| {\textbf{Confounder Approach}}  \\
        |[block]| {Under treatment effect heterogeneity, is the LATE useful?}    &  \\
        |[block]| {Sample size and proportion of population that receives treatment}    &  
        |[decision]| {Low sample size + weak IVs and/or extreme proportion} \\
        |[decision]| {Low sample size + OK proportion}  & |[decision]| {High sample size + OK proportion}  \\ 
        |[block]| {\textbf{IV Approach:} 2SLS with strong IVs} & 
        |[block]| {\textbf{IV Approach:} 2SLS or ML methods with weak or strong IVs}\\
    };
    \path [>=latex,->] (m-1-1) edge (m-2-1);
    
    \path [>=latex,->] (m-2-1) edge node[left]{Yes} (m-3-1);
    \path [>=latex,->] (m-2-1) edge node[above]{No} (m-3-2);
    
    \path [>=latex,->] (m-3-1) edge node[left]{Yes} (m-4-1);
    \path [>=latex,->] (m-3-1) edge node[above]{No} (m-3-2);
    
    \path [>=latex,->] (m-4-1) edge node[left]{Yes} (m-5-1);
    \path [>=latex,->] (m-4-1) edge node[above]{No} (m-3-2);
    
    \path [>=latex,->] (m-5-1) edge (m-5-2);
    \path [>=latex,->] (m-5-2) edge (m-3-2);
    
    \path [>=latex,->] (m-5-1) edge (m-6-1);
    \path [>=latex,->] (m-5-1) edge (m-6-2);
    
    \path [>=latex,->] (m-6-2) edge (m-7-2);
    \path [>=latex,->] (m-6-1) edge (m-7-1);
\end{tikzpicture}
\end{center}
\end{figure}

In this section, we offer general principles and guidance in a series of steps to help an analyst weigh the validity of the confounder approach to the IV approach. Why not utilize both approaches in the same analysis? Certainly, if the appropriate variables are available then one can use IV estimates as a sensitivity analysis of confounder adjustment estimates and vice versa. An issue, however, arises when each approach offers conflicting results such as oppositely signed effect estimates or when one is statistically significant and not the other. In these cases, one will have to judge which estimate is more desirable in terms of robustness regarding consistency, utility, and generalizability. A visualization of the heuristic in this section is presented in Figure \ref{fig:method selection}.

The first step is to assess the severity of unobserved confounding. In our view, compared to a confounder adjustment approach, the IV approach trades increased complexity and decreased interpretability for potential consistency in the setting of unmeasured confounding. Therefore, unless there is evidence to suggest that the confounder approach will suffer from notable unobserved confounding, the more straightforward approach should be pursued. In order to evaluate unobserved confounding, one may identify confounders separate from the dataset and then assess the dataset at hand for what is present, what is missing, and what observed variables could serve as proxies for missing confounders. Furthermore, one should postulate the magnitude of unobserved confounding as there could be scenarios where the lion's share of variation due to confounding can be captured by a few observed confounders. For example, a rich EHR dataset may not be missing important confounders of interest whereas an insurance claims dataset is frequently missing valuable medical information. In addition, because the IV approach largely builds upon the confounder approach by incorporating confounders in the first and second stages, identifying confounders remains important to help gauge the validity of potential IVs.

When performing this step, one should keep in mind how conditioning on certain confounders may affect interpretability and reproducibility. Further, one may consider the potential for treatment effect heterogeneity as this also affects the generalizability of confounder methods via effect modification, keeping in mind that omitted confounders could also be effect modifiers.\citep{vander_weele_confounding_2012} In reviewing treatment effect heterogeneity, it also may be a good time to weigh the usefulness of marginal versus conditional treatment effects for the purposes of the study at hand. This will assist in interpreting regression adjustment and IPTW output.

In the second step, one should identify whether there are any potential IVs in the paradigm and then in the dataset. Even if a confounder approach will be used, adjusting for IVs can amplify inconsistency.  Theoretically and empirically, one should explore the strength of the IVs. From this, we can hypothesize the degree of invalidity of the selected IVs in terms of the independence assumption. The order is important: if the IVs are strong but slightly invalid then the degree of inconsistency can still be better than a confounder approach; if the IVs are seemingly weak but valid then they can potentially be salvaged by more sophisticated algorithms. Just how invalid an IV can be to still be better than confounder methods is a developing topic of research. Of note, Pearl (2010) investigated this question with one confounder and one imperfect IV using a threshold of when to choose the candidate variable as a confounder.\citep{pearl_class_2010} Nevertheless, the interpretability of this threshold could be improved to better accommodate applied analyses and does not consider the potential results from an IV estimation procedure such as 2SLS.

If there is notable treatment effect heterogeneity per the previous step, one should describe the subpopulation characterized by the LATE and any potential differences between the LATE and ATE. One should also justify whether this subpopulation is worth studying in the first place. If the subpopulation is too obscure, a biased confounder approach is perhaps of more scientific value.

The third and final step is to examine other relevant characteristics of the dataset: generally, the sample size and, if the treatment is binary, the distribution of the proportion of individuals that received the treatment. Large sample sizes cannot salvage confounder-based estimates in the presence of a large amount of unobserved confounding but can greatly help the IV approach. Notably, IV methods have larger standard errors for the treatment effect compared to a confounder method like OLS, which is only exacerbated by weak IVs. In some cases, one could argue that IV methods could produce Type II error rates so large that the analysis is not worth pursuing. Therefore, not only will a larger sample size create a more precise estimate and lessen finite estimation bias but, in addition, the sample size will open the door to more sophisticated methods to be used reliably like ML and cross-fitting. Accordingly, we may strengthen the first stage and decrease sensitivity to exclusion restriction violations.

For a binary treatment, the distribution of the proportion of the population that receives the treatment is important both for the validity of the positivity assumption for propensity score-based methods and for the validity of using an LPM or 2SRI for the IV first stage. Probabilities near 0 or 1 are an issue for both approaches but it remains to be investigated whether positivity violations cause more inconsistency in confounder methods when compared to 2SLS. Under near-positivity violations, we may also run into concerns surrounding Type II error as we would for IV methods. We note these notions are particular to the scenario but sample size does not necessarily mitigate the effects of near-positivity violations under extreme imbalance of treatment assignment.

A notable data-driven diagnostic of the relative performance of confounder adjustment methods versus IV methods under observed confounding (recall that unobserved confounding can also violate the independence of the assumption of an IV) is the bias ratio.\citep{jackson_toward_2015} Suppose we have the OLS in Eq. \ref{simple_ols}:

\begin{equation}\label{simple_ols}
    E[(y_{i,t=2} - y_{i,t=0})|D,X] = \beta_0 + \beta_1 d_i + \beta_2x_{1i}.
\end{equation}

If we have a binary IV, we can derive the bias from omitting the confounder $X_1$ for OLS and 2SLS as
\begin{equation}\label{eq:bias_ols}
    Bias_{OLS}(X_1) = \beta_2(E[X_1|D=1]-E[X_1|D=0]),
\end{equation}
\begin{equation}\label{eq:bias_2sls}
    Bias_{2SLS}(X_1) = \beta_2\frac{E[X_1|Z=1]-E[X_1|Z=0]}{E[D|Z=1]-E[D|Z=0]}.
\end{equation}

We can take the ratio of Eq. \ref{eq:bias_ols} and Eq. \ref{eq:bias_2sls} to obtain a bias ratio where a value greater than 1 implies that the 2SLS estimate is more sensitive to an omitted confounder than the OLS. If we assume that unobserved confounding is correlated with the observed confounder that we intentionally omit then we gain information about each method's sensitivity to unobserved confounding. Improvements include adding variance to the bias ratio to represent uncertainty in estimates and considering the magnitude of $\beta_2$ in our ratio.\citep{davies_how_2017,zhao_graphical_2018}.

\section{Illustrative Example: Off-label Donepezil for MCI patients}

We now turn to our applied data analysis surrounding donepezil, a cholinesterase inhibitor, off-label use in MCI patients. We will pursue the causal inference analysis steps outlined throughout the paper: first, we will outline the scientific paradigm of the analysis and define confounders and instruments; next, we will examine our available dataset and its ability to answer the causal question of interest and limitations on this goal; lastly, we will compare and contrast confounder and IV methodologies with respect to causal effect estimates.

We analyze data from the Alzheimer's Disease Neuroimaging Initiative (ADNI), an observational dataset that longitudinally tracks cognitively normal, MCI, and AD volunteers aged 55-90 with good general health. Among other information, the data includes both general health and specific neurocognitive data collected roughly every six months until the patient leaves the study.\citep{mueller_ways_2005} Our dataset contains patients from several ADNI recruitment waves beginning in October 2004 until May 2021 when the data was extracted for analysis. Of note, those already on cholinesterase inhibitors such as donepezil are allowed into ADNI if they have been stable on the medication for at least 12 weeks prior to entry.

\subsection{Study Paradigm}

Our goal is to estimate the difference between donepezil users' and non-users' two-year change from baseline in ADAS-cog. Furthermore, we are potentially susceptible to confounding by indication, which warrants using the confounder and IV approaches to isolate the causal estimand of interest. 

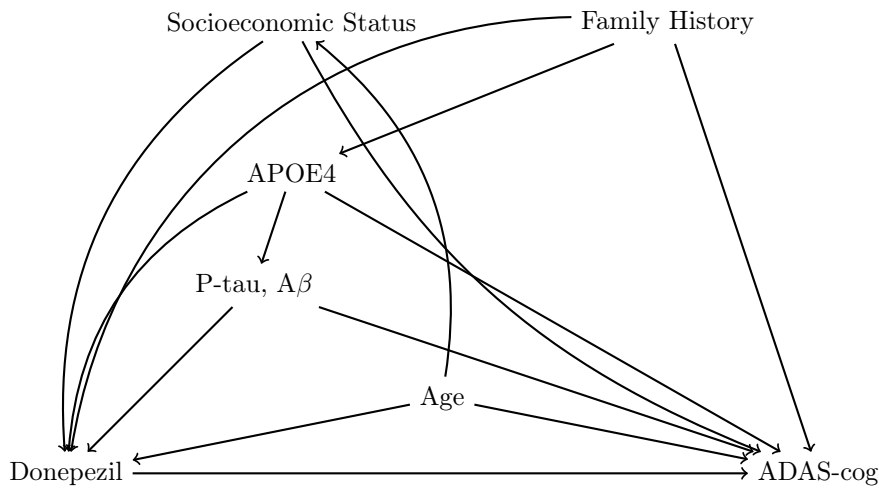
\begin{figure}
        \centering
    \begin{tikzpicture}[thick]
        \node (1) at (0,0) {Donepezil};
        \node (2) at (10,0) {ADAS-cog};
        \node (4) at (5,1) {Age};
        \node (5) at (3,6) {Socioeconomic Status};
        \node (6) at (3,4) {APOE4};
        \node (7) at (8,6) {Family History};
        \node (8) at (2.5,2.5) {P-tau, A$\beta$};
        
        \path [->] (1) edge (2);
        
        \path [->] (4) edge (1);
        \path [->] (4) edge (2);
        \path [->] (4) edge[bend right=30] (5);
        
        \path [->] (5) edge[bend right=30] (1);
        \path [->] (5) edge [bend right=20](2);

        \path [->] (6) edge[bend right=30] (1);
        \path [->] (6) edge (2);
        
        \path [->] (7) edge[bend right=40] (1);
        \path [->] (7) edge (2);
        \path [->] (7) edge (6);
        
        \path [->] (8) edge (1);
        \path [->] (8) edge (2);
        \path [->] (6) edge (8);
    \end{tikzpicture}
    \caption{Directed Acylclic Graph of Donepezil Analysis with a Notable Confounders}
    \label{fig:ADNI_dag}
\end{figure}

To identify confounders, we begin by listing potential general predictors of ADAS-cog scores such as age, sex, family history of neurodegenerative disease, socioeconomic status, physical disabilities, race, and comorbidities. In addition, we can list particular biological attributes: the number of APOE4 alleles and cerebral spinal fluid (CSF) measurements of phosphorylated tau (P-tau) and amyloid-$\beta$ (A$\beta$). Except for physical disabilities, all of the aforementioned variables could be potential influences on the assignment of donepezil and, hence, we consider them confounders. A hypothesized DAG of causal relationships with a few of the potential confounders is presented (Figure \ref{fig:ADNI_dag}).

If there are differential prescriber attitudes to donepezil then physician and facility prescribing patterns could be used as IVs.\citep{chen_use_2011} A similar possible IV is the time since Food and Drug Administration (FDA) approval of donepezil for AD patients in 2004. In the roughly two decades since approval, a large volume of research has been conducted on the MCI construct and donepezil in MCI patients. As a result, the general perspectives of practitioners may have changed over time. Of course, we must assume the IV is monotonic; that is, attitudes towards donepezil use are monotonically more favorable or less favorable over time.

Another potential category of IVs is related to contraindications such as bradycardia or gastrointestinal disorder, which may make off-label prescription unfavorable relative to benefits. Importantly, we should note that contraindication IVs are sometimes related to the general health of a patient, which is, in turn, related to cognitive decline. In this case, we should be particularly cautious in utilizing certain contraindications as IVs and should only do so if we can adjust for most of the related confounders.

Identifying possible IVs allows us to examine whether the LATE is of scientific interest. All of the IVs previously mentioned essentially capture variation caused by differential physician prescribing attitudes. Therefore, under these IVs, a complier is a patient that follows the judgments of the physician and the LATE would be this population's treatment effect. The resulting estimate for the complier treatment effect could be useful because a formal analysis using this LATE to evaluate donepezil off-label practice intends to influence guidelines for physicians, which are constructed with the assumption of patient compliance to physician advice. Defiers for these IVs are unlikely as individuals usually follow physicians' judgments regarding risks and benefits.

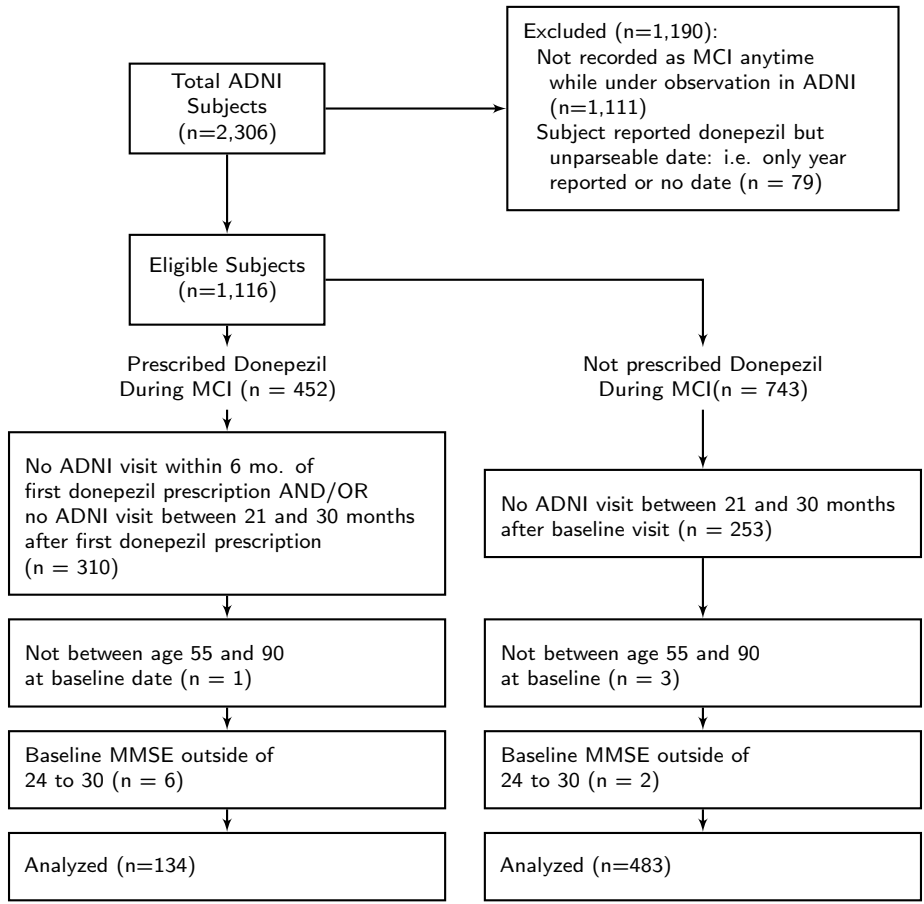
\begin{figure}[h!]
    
    \begin{center}
  \captionof{figure}{Flowchart of selection of final cohort}
  \sffamily
  \footnotesize
  \label{fig:CONSORT}
  \begin{tikzpicture}[auto,
    block_center/.style ={rectangle, draw=black, thick, fill=white,
      text width=8em, text centered,
      minimum height=4em},
    block_left/.style ={rectangle, draw=black, thick, fill=white,
      text width=16em, text ragged, minimum height=4em, inner sep=6pt},
    block_noborder/.style ={rectangle, draw=none, thick, fill=none,
      text width=18em, text centered, minimum height=1em},
    block_assign/.style ={rectangle, draw=black, thick, fill=white,
      text width=18em, text ragged, minimum height=3em, inner sep=6pt},
    block_lost/.style ={rectangle, draw=black, thick, fill=white,
      text width=16em, text ragged, minimum height=3em, inner sep=6pt},
      line/.style ={draw, thick, -latex', shorten >=0pt}]
    \matrix [column sep=5mm,row sep=3mm] {
      \node [block_center] (referred) {Total ADNI \\ Subjects (n=2,306)};
      & \node [block_left] (excluded1) {Excluded (n=1,190): \\
        \h Not recorded as MCI anytime \\ 
        \hh while under observation in ADNI \\
        \hh (n=1,111) \\
        \h Subject reported donepezil but \\
        \hh unparseable date: i.e. only year \\
        \hh reported or no date (n = 79)};\\
      \node [block_center] (random) {Eligible Subjects\\(n=1,116)}; 
      & \\
      \node [block_noborder] (i) {Prescribed Donepezil \\ During MCI (n = 452)}; 
      & \node [block_noborder] (wlc) {Not prescribed Donepezil \\ During MCI(n = 743)}; \\
      \node [block_assign] (i_visit) { \\No ADNI visit within 6 mo. of \\
      first donepezil prescription AND/OR \\
      no ADNI visit between 21 and 30 months\\
      after first donepezil prescription \\
      (n = 310)}; 
	  & \node [block_assign] (c_visit) {\\No ADNI visit between 21 and 30 months \\
      after baseline visit (n = 253)}; \\
      \node [block_assign] (i_age) {\\Not between age 55 and 90 \\
      at baseline date (n = 1)};
	  & \node [block_assign] (c_age) {\\Not between age 55 and 90 \\
      at baseline (n = 3)}; \\
      \node [block_assign] (i_mmse) {Baseline MMSE outside of\\ 
      24 to 30 (n = 6)};
	  & \node [block_assign] (c_mmse) {Baseline MMSE outside of\\ 
      24 to 30 (n = 2)}; \\
      \node [block_assign] (i_ana) {Analyzed (n=134)};
      & \node [block_assign] (c_ana) {Analyzed (n=483)}; \\
    };
    \begin{scope}[every path/.style=line]
      \path (referred)   -- (excluded1);
      \path (referred)   -- (random);
      \path (random)     -- (i);
      \path (random)     -| (wlc);
      \path (i) -- (i_visit);
      \path (i_visit) -- (i_age);
      \path (i_age) -- (i_mmse);
      \path (i_mmse) -- (i_ana);
      \path (wlc) -- (c_visit);
      \path (c_visit) -- (c_age);
      \path (c_age) -- (c_mmse);
      \path (c_mmse) -- (c_ana);
    \end{scope}
  \end{tikzpicture}
\end{center}
\end{figure}

\subsection{Cohort Selection}

A flowchart of the selection criteria is depicted visually in Figure \ref{fig:CONSORT}. We ascertained off-label prescription donepezil from self-reported pharmaceutical use and commencement dates collected at every visit in ADNI. Using keyword search, we first found patients that were prescribed donepezil and then used diagnosis data to identify prescriptions during MCI. Patients who had never reported use of donepezil were considered to be a member of the control group. Furthermore, individuals who received donepezil while having AD were also used as controls but their data was artificially censored at the visit before the first donepezil prescription.

For the treatment group, the start date was the self-reported commencement of donepezil and we required that the baseline measurement of ADAS-cog must be from the closest visit within six months of this start date. If no such visit existed, these subjects were dropped from the study. Under an intent-to-treat principle,\citep{gupta_intention--treat_2011} if we were able to observe a patient beginning donepezil, then they remained in the treatment group. The end date was computed by adding two years to the start date and taking the closest visit between 21 months to 30 months after the start date. Once again, if no such visit existed then the subjects were dropped. 

The control group start date was the first visit after MCI diagnosis. For many, this was their ADNI entry visit. Like the treatment group, the end of the date was computed by adding two years to the start date and applying the same 21 to 30 month requirement.

Mimicking the Peterson et al 2005 trial, on top of the ADNI inclusion criteria, we applied further inclusion-exclusion criteria from this trial depending on the availability in the data. Specifically, we only included those with an MMSE of 24-30 at baseline and excluded anyone with self-reported bipolarism, schizophrenia, suicide ideation, or psychosis. Applying these criteria resulted in 134 donepezil subjects and 483 control subjects.

\subsection{Examining the Data: Confounders and IVs}

Considering Figure \ref{fig:ADNI_dag}, which depicts major confounders, we hypothesize that there are notable unobserved confounders. For example, we cannot adjust for socioeconomic status (SES) nor family history. SES is not in the dataset while family history contains too many missing values to impute (approximately 93\% before inclusion criteria). We have information on comorbidities via three indicator variables: the presence of cardiovascular disease, other neurological disease, and renal disease. These indicators are limited as they fail to capture severity and not all the conditions these categories include are relevant. Overall, the impact of these missing or low-quality confounders is not too large with the exception of family history, which is almost certainly asked during a doctor's visit. 

Another potential area of unobserved confounding could come from the lack of historical information on a patient: we do not have neurocognitive test scores on these patients before entry into ADNI but a prescriber may have this information and consider it before making a decision. Most subjects' study baseline visits are either close or at ADNI entry, which precludes the calculation of any sort of useful measure of history from the data such as a first-order trend. In addition, there are many difficult to measure evaluations of patient well-being that stem from the intuition of the attending clinician.

For observed variables that may serve as confounders, we have demographic information on sex, age, years of education, and APOE4 status. One necessary confounder induced by our sampling scheme is time length, in years, the patient was in ADNI before the start of the analysis. This is associated with the outcome because the longer someone has MCI, the more opportunity there is for cognitive decline. Moreover, time in ADNI is associated with treatment assignment as most of the start dates for the control group were entry into ADNI while the start dates of the donepezil group tended to be several months into ADNI (Table \ref{tab:descriptives}).

We utilize CSF measurements for P-Tau and A$\beta$ but only at ADNI entry because missingness was too high otherwise (36.3\% for both variables' entry measurements and 84.3\% for all recorded measurements). We believe that for individuals with start dates further from baseline, combining entry P-Tau and A$\beta$ measurements with time in ADNI is sufficient to extrapolate to the study baseline.

\begin{table}[]
    \begin{center}
    \captionof{table}{Potential confounders stratified by donepezil use} \label{tab:descriptives} 
    \begin{tabular}{ p{2in} p{1in} p{1in} p{1in}}\toprule[1pt]
    \bf & \bf Donepezil (n=134) & \bf Control (n=483) & \bf SMD$^a$ \\\midrule
    Cardiovascular Issues – Yes (n (\%))$^{c}$  &  96 (71.6) & 342 (70.8) & 0.018 \\
    Neurological Issues – Yes (n (\%))$^{c}$  &  43 (32.1) & 171 (35.4)  & 0.070\\
    Renal Issues - Yes (n (\%))$^{c}$  &  67 (50.0) & 210 (43.5)  & 0.131\\
    Age (mean (SD))  &  74.92 (6.69) & 72.96 (7.71) & 0.271\\
    Sex – Male (n (\%))  &  81 (60.4) & 282 (58.4) & 0.042\\
    Years Education (mean (SD)) &  15.81 (2.65) & 16.23 (2.67) & 0.158\\
    APOE4 Count (n (\%)) & & & 0.306\\
    \-\hspace{0.1in}0   &  60 (44.8) & 283 (58.6)&  \\
    \-\hspace{0.1in}1   &  56 (41.8) &  166 (34.4)&  \\
    \-\hspace{0.1in}2   & 18 (13.4) & 34 (7.0) &  \\
    Time in ADNI (mean (SD))  &  1.19 (2.03) & 0.41 (1.50)  & 0.437\\
    A$\beta$ (mean (SD))$^{b,c}$  &  781.21 (354.35) & 1096.50 (441.29)  & 0.788\\
    P-tau (mean (SD))$^{b,c}$   &  29.61 (12.61) & 26.17 (12.63)  & 0.272\\
    Starting ADAS-cog (mean (SD))$^{c}$  &  11.90 (4.46) & 8.88 (3.96)  & 0.714\\
    \bottomrule[1.25pt]
    \end {tabular}\par
    \end{center}
    \-\hspace{0.42in}\footnotesize{$^a$Standardized mean difference across groups}\\
    \-\hspace{0.42in}\footnotesize{$^b$Measured at ADNI entry – may not coincide with their baseline start date}\\
    \-\hspace{0.42in}\footnotesize{$^c$Includes imputed values from one of the five imputed datasets}\\
\end{table}

Overall, while the observed confounders serve as useful proxies of the unmeasured components, we judge that there still remains unobserved confounding by indication. Therefore, it is warranted to investigate the IV approach. Unfortunately, ADNI did not lend itself to intuitive and high-quality IVs. Nevertheless, we utilized time, in years, since FDA approval of donepezil and contraindications in the form of indicator variables of encompassing cardiovascular issues such as bradycardia and sick sinus syndrome, gastrointestinal disorder, and asthma. More details about these IVs including investigations of strength and validity can be found in the supplementary appendix. Notably, these IVs were weak but the LATE is scientifically meaningful for reasons discussed earlier.

We note that the weakness of the IVs combined with a lower sample size cautions us against using IVs methods. However, for the purposes of comparison to confounder methods, we will still fit models with these IVs. Lastly, the proportion of individuals in the dataset who took the treatment is not extreme so the use of the LPM in the IV methods is valid.

\subsection{Statistical Methodology}

Before applying selection criteria, ADAS-cog, MMSE, A$\beta$ at ADNI baseline, and P-tau at ADNI baseline were missing for roughly 30\% of the observations. The comorbidity and contraindication (IV) indicators were also missing but only for 0.4\% of the observations. These values were imputed using multiple imputation via chained equations (MICE) resulting in five different datasets. Specifically, to account for within-subject correlation, we performed predictive mean matching using a linear mixed model with the fixed effects of time in ADNI, clinical dementia rating scale sum of boxes, MMSE, ADAS-cog, age at ADNI entry, sex, APOE4 count, whether they had taken donepezil, the comorbidity indicators, $\beta$ at ADNI baseline, and P-tau at ADNI baseline. We included a random intercept and a random slope on time in ADNI.

To obtain a "baseline" estimate of the treatment effect without any adjustment of confounders, we regressed the change in ADAS-cog on treatment assignment. Then, we utilized the variables in Table 3 with the exception of starting ADAS-cog as confounders for OLS, IPTW, TMLE, and DML. For IPTW, the propensity score was computed using logistic regression. Diagnostics are reported in the appendix. For TMLE, we utilized five-fold cross-fitting (5) and random forests to fit nuisance quantities. For DML, random forests were also used for fitting models involving nuisance quantities. The IV analyses included the same confounders and the aforementioned IVs. For models, we fit 2SLS, PL regularizing the first-stage, and DML that used random forests for nuisance quantities. To compare the strength of the first-stage, pooled $F$-statistics were computed on the first-stage OLS involving all the IVs and confounders and the IVs and confounders selected by PL. Estimates across imputed datasets were combined using Rubin's rules. All analysis was conducted using R Version 4.0.3.

\subsection{Results}

\begin{center}
\captionof{table}{Treatment effect estimate by methodology} \label{tab:ADNI_results} 
\begin{tabular}{ p{2in} p{1in} p{1in} p{0.75in}}\toprule[1pt]
\bf Method & \bf Estimate & \bf Std. Err. & \bf P-value \\\midrule
Unadjusted OLS  &  2.12 & 0.58 & <.001  \\
OLS  &  1.38 & 0.62 & 0.029  \\
IPTW  &  1.46 & 0.99 & 0.157 \\
TMLE  &  1.48 & 0.43 & <0.001 \\
DML 1 – Confounders Only  &  2.62 & 1.56 & 0.092 \\
2SLS & -0.09 & 2.85 & 0.97\\
Post-LASSO & -0.56 & 2.89 & 0.85\\
DML 2 - Confounders and IVs & -2.00 & 4.22 & 0.634\\
\bottomrule[1.25pt]
\end {tabular}\par
\end{center}

The results, presented in Table \ref{tab:ADNI_results}, demonstrate a clear divide between the confounder and IV approaches. While the confounder methods found a point estimate that suggests donepezil has a deleterious effect on MCI patients, the IV methods produce an attenuated but beneficial effect estimate. Taking into account the standard errors, however, the IV methods cannot rule out a null effect of donepezil.

Among the confounder methods, differing levels of efficiency led to some methods attaining statistical significance using a level .05 test, while others did not (with the exception of the unadjusted OLS). TMLE yielded the lowest estimated standard error. In contrast, DML 1 with random forests had a large standard error possibly due to the lower sample size and less efficient use of the data in cross-fitting. When we used DML 1 with logistic regression and OLS to estimate nuisance quantities, the point estimate and standard error of the estimate decreased to be closer with the other confounder methods (Estimate: 1.48, SE: 1.16). Similarly, substituting linear models for random forests in TMLE increased the standard errors but did not modify the point estimate though it was not significant (Estimate: 1.33, SE: 0.99). Subsequently, the estimate became statistically insignificant ($p=0.178$). OLS and IPTW had slightly different results with IPTW having a larger point estimate (i.e. a more harmful effect) and a larger standard error. The different point estimates could be due to slight treatment effect heterogeneity while the lower precision of IPTW estimate could be due to near positivity violations (see Appendix).

\begin{figure}[!htb]
    \centering
    \caption{Post LASSO Estimates and 95\% Confidence Intervals by Shrinkage Amount ($\lambda$)}\par\medskip
    \includegraphics[scale=0.25]{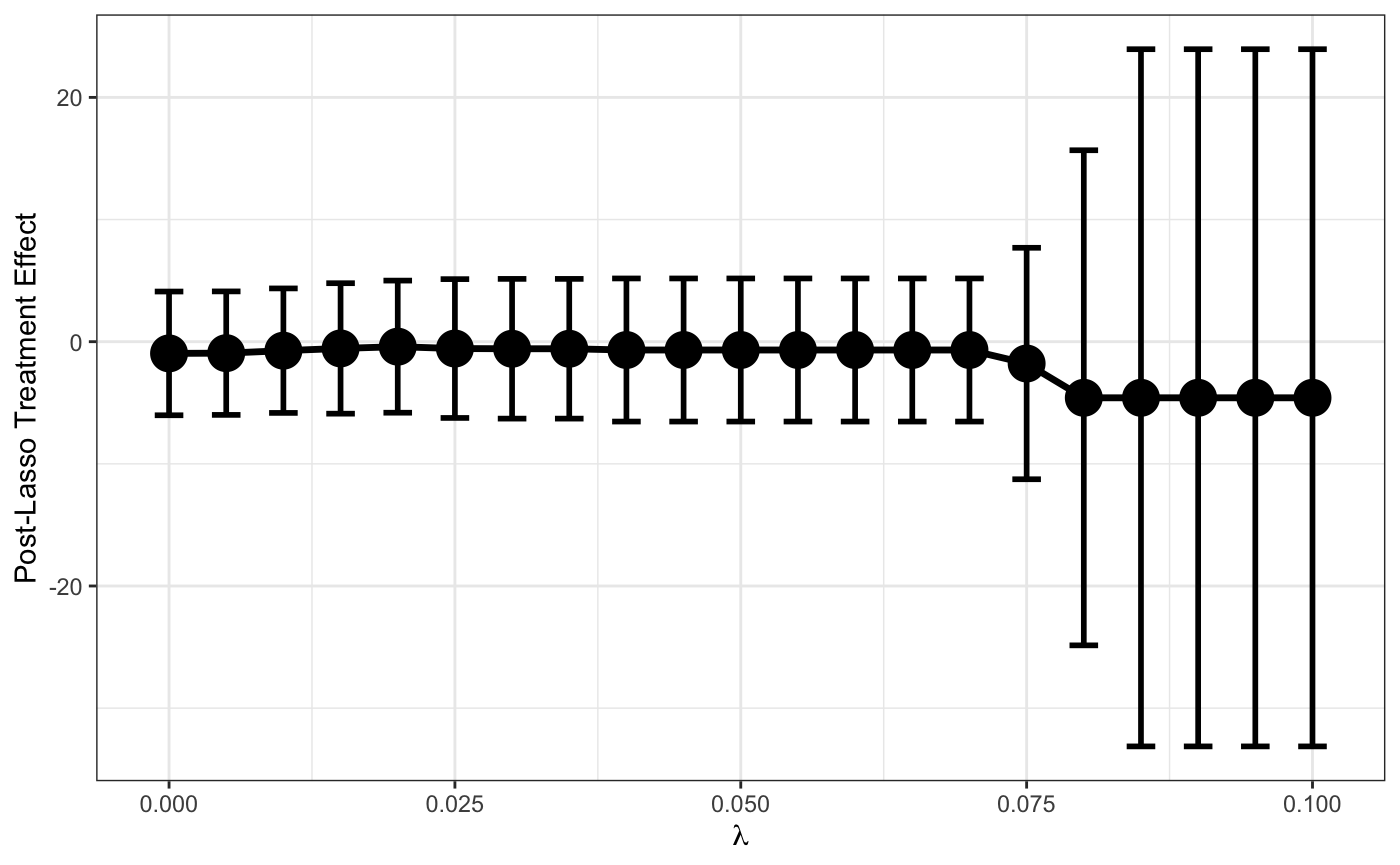}
    \label{fig:PL_results}
\end{figure}

Regarding PL, we originally intended to utilize the authors' R package (named \texttt{hdm}) default "data-driven" selection for $\lambda$, the regularization parameter in the first stage LASSO.\citep{chernozhukov_high-dimensional_2016} However, the package-selected  value of $\lambda$ led to all IVs being eliminated from the first stage and subsequently unreliable results. Figure \ref{fig:PL_results} shows that at $\lambda = 0.075$, when the instruments were all excluded, the point estimate and standard errors change notably. Noting that when there was at least one IV the results largely remained the same, we reported the results from $\lambda = 0.025$.

Across all imputed datasets, the first-stage LASSO with $\lambda = 0.025$ selected time in ADNI, baseline A$\beta$, and time since FDA approval. For two imputed datasets, LASSO retained baseline P-tau and, in one of these two datasets, asthma was additionally retained. Selecting variables strengthened the first-stage with the PL first-stage fit being much higher than 2SLS (pooled $F = 33.3$ and pooled $F = 8.5$, respectively).

Ostensibly because DML 2 builds upon PL with better practices and more sophisticated ML techniques, we would expect similar results but this was not the case. The point estimate was the most negative and most extreme besides the unadjusted OLS. One explanation could be related to Figure \ref{fig:PL_results} where "over-regularization" of the first-stage produces extreme results. When we replaced random forests for OLS in DML 2 we obtained results more similar to the other IV methods (Estimate: 0.154, SE: 2.712) suggesting a sensitivity of DML point estimates to the choice of the prediction model in both approaches. We were not able to fit DML with LASSO as the authors' official package (\texttt{dml}) did not support non-numeric covariates specifically for the LASSO algorithm.

\subsection{Takeaways from Applied Example}

In this analysis, we found that the two different approaches presented two different conclusions. Contextualizing these results, the Peterson et al. clinical trial found a -0.27 point change in ADAS-cog after 2 years for the donepezil group compared to placebo that failed to attain statistical significance (95\% CI: -1.13 to 0.59 points),\citep{petersen_vitamin_2005} which was similar to the IV estimates. Meanwhile, our confounder-based estimates mirror a similar analysis using ADNI by Schneider, Insel, and Weiner conducted in 2011.\citep{schneider_treatment_2011} Using a linear mixed model, the authors found a statistically significant increase in the rate of decline in donepezil users compared to control subjects. Using this slope to calculate the difference from baseline after two years, we arrive at a comparable estimate of 1.56.

Adding the more sophisticated ML-based algorithms did not show notable changes over traditional methods and even created a few dilemmas. For PL, the default algorithm only included baseline A$\beta$, selecting out all IVs likely due to their weakness. This was "a feature, not a bug": the first-stage $F$-statistic was indeed the highest at $60.94$, indicating a "stronger" first-stage. But without an IV in the first-stage, the results are theoretically meaningless, which cautions against a purely "greedy" approach to fitting the first stage.

Another issue with PL was that across imputed datasets, the variables selected for the first stages were different. Thus, both the confounder and IV estimands across imputed datasets are not exactly the same. For example, the estimated LATE from selecting only time since FDA approval as IV was combined with the estimated LATE using both time since FDA approval and asthma. If this is the case, Rubin's rules pooled estimates are actually an average of different LATE, which is ambiguous. This is yet another problem with a "greedy" approach to the first-stage as it is likely that a similar process occurred for TMLE and DML random forests.

One interesting discussion surrounds the choice of hyperparameters for TMLE and DML, particularly which ML algorithms should be selected given that the results changed depending on the algorithm chosen: TMLE-derived results became insignificant while the DML 2 point estimate flipped signs. This raises concerns about replicability issues that may arise from an arbitrary choice of ML algorithm. The criteria for selecting ML algorithms is presently unclear from both the authors of TMLE and DML. In our case, the choice of random forest was made before any results were computed.

This analysis contains several limitations. Firstly, the selection criteria were not exactly the same as the Peterson et al and Schneider et al, leading to the populations we were estimating the effect of donepezil to be slightly different. The exclusion of MCI individuals from ADNI who have not been stable on donepezil skews the results towards the null as these individuals likely experienced harm from donepezil but were missing from the data. In addition, the control group in our analysis is aware they are not getting donepezil whereas in a trial, the control group received a placebo in a blinded fashion. Selecting the baseline and after 2-year measurement dates could use improvement where, instead, we could have used longitudinal modeling and modeled the rate of decline in ADAS-cog. Unfortunately, the more sophisticated ML methods are not yet equipped to handle clustered data. Even so, the main purpose was to compare our results to Peterson et al, which utilized 24-month change. Starting and ending dates often were missing the day of the month leading to inaccuracy in measuring exact 24 month changes. For the instrumental variable strategy, we were relying on self-reported health issues, a reporting process that suffers from recall bias. Most likely there were omitted or misdiagnosed conditions, which meant that our IVs were much weaker than they could have been given accurate medical history. Lastly, dosage of donepezil and compliance could be two critical nuances in evaluating treatment effects that we were not able to measure here. As an example of compliance, it could be that the already more severe MCI subjects would also have issues remembering to take their medication. Despite these limitations, the variation in estimands and resulting estimates between confounder adjustment and IV approaches remains of interest.

\section{Summary and Discussion}

In this paper, we have outlined a paradigm for isolating causal effects in observational studies that centers around two approaches: confounders and IVs. We began by defining each approach and its often untestable assumptions. We then discussed a set of heuristics for executing the approaches from identifying relevant variables to examining external validity. In addition, we highlighted that the two approaches often overlap (e.g. potential bias amplification by IVs used as confounders) and, thus, that one approach cannot be undertaken without some consideration of the other. Following this, we discussed three popular methodologies, OLS, IPTW, and 2SLS, for conducting statistical modeling along the confounder and IV approach including comparing OLS to IPTW and why causal inference modeling should not simply be viewed as a prediction task. Further, we explored the implications of violations in the assumptions of each of these methodologies and how to operate when these assumptions could be violated. After touching upon causal inference in non-linear scenarios and more flexible models, we outlined a set of steps to consider when weighing each approach.

The principles and processes laid out in the above portions culminated in the applied example surrounding off-label donepezil where each approach provided different challenges and, ultimately, conflicting estimates. A natural question to ask is: which approach was better at isolating the causal effect of interest? Of course, we do not have knowledge of the true causal effect in the general population, so it is impossible to have a definitive answer to this question. This mostly owes to the untestable assumptions discussed throughout the paper such as proving we have accounted for virtually all confounding and that we have valid IVs. Even then, the confounder and IV estimates may be targeting different causal estimands and, furthermore, the process of outlining the scientific paradigm and causal mechanisms (e.g. identifying IVs) included many conjectures. Nevertheless, for sake of argument, we can speculate on the above question by judging potential scenarios in our analysis and how they may impact each approach.

In surmising what the true ADNI causal effect could be, we can first consider the Peterson et al trial as we know this estimate is, on average, unbiased for the treatment effect in the trial population. But are the populations and usage of donepezil in ADNI at the time we obtained the data similar to that of the RCT? On one hand, recruitment criteria for ADNI are akin to MCI clinical trials and we imposed extra inclusion-exclusion criteria similar to Peterson et al. On the other hand, there may be cohort effects from 2005 to the present as the understanding of MCI has developed since then and, further, we cannot fully account for compliance and concomitant medicines. 

Certainly, if we assumed the clinical trial was a good benchmark for our study, the IV methodology would be the better performer. However, we must weigh the fact that the lack of high-quality IVs and low sample size likely led to large finite sample bias and standard errors. The difficulty in finding IVs underscores the fact that confounders are much more intuitive to source than IVs. Yet, both are equally important as paths to consistent treatment effects, and, as such, one aim of this paper is to increase awareness and education of IVs such that they can be identified and investigated as easily as confounders.

Considering the available observational study evidence, particularly Schneider, Insel, and Weiner's ADNI study, we can pose two questions. Firstly, like in the previous paragraph, how different is the population of our study? Secondly, seeing that our confounder results are similar, can we postulate sources of unobserved confounding and hypothesize how these sources have skewed our results? To answer the first question, because we have more present data, we can certainly re-perform our analysis on the subset of ADNI that Schneider had (they pulled their data on May 2009) to observe how the results changed. Doing this, we observed similar observational study estimates. But this far from disproves that there is no cohort effect that makes our populations irreconcilable. 

For the second question, one possible explanation for confounding by indication is that more severe patients receive donepezil, which can skew a confounder estimate that fails to fully account for this towards one that suggests harm. Although, we are limited in this line of thinking because of the difficulty of making accurate statements regarding how a physician decides if a patient is "severe enough" to be prescribed donepezil. If we were to suppose the doctor looks at the steepness of the rate of decline then a historical measure such as a first-order trend of their cognitive tests scores could mitigate some confounding. This highlights the need for more research to specifically quantify factors contributing to physician decision making in specific disease areas. 

To conclude, our goals with this paper were, firstly, to highlight the key differences between the confounder and IV approaches and estimates, and, secondly, to lay out a set of principles and heuristics that can help an analyst choose the best tool under the potential violation of untestable causal assumptions. The key is to favor the approach that one has judged to violate the assumptions to a lesser degree than the other. Of course, if there are assumptions we can do nothing about, one could decide to simply not pursue causal estimates from observational studies. Yet, this would discount the compelling scientific reasons to conduct observational studies in health research such as for hypothesis generation. Therefore, skeptical yet practical discussions of how to best move forward and avenues of further development, such as the ones in the paper, are of utmost necessity.

\section{Acknowledgements and Disclosures} 

RSZ and DLG are supported by NIH/NIA P30AG066519 and NIH/NIA RF1AG075107. JDG is supported by NIH/NIA AG066519.

Data collection and sharing for this project was funded by the Alzheimer's Disease
Neuroimaging Initiative (ADNI) (National Institutes of Health Grant U01 AG024904) and
DOD ADNI (Department of Defense award number W81XWH-12-2-0012). ADNI is funded
by the National Institute on Aging, the National Institute of Biomedical Imaging and
Bioengineering, and through generous contributions from the following: AbbVie, Alzheimer’s
Association; Alzheimer’s Drug Discovery Foundation; Araclon Biotech; BioClinica, Inc.;
Biogen; Bristol-Myers Squibb Company; CereSpir, Inc.; Cogstate; Eisai Inc.; Elan
Pharmaceuticals, Inc.; Eli Lilly and Company; EuroImmun; F. Hoffmann-La Roche Ltd and
its affiliated company Genentech, Inc.; Fujirebio; GE Healthcare; IXICO Ltd.; Janssen
Alzheimer Immunotherapy Research \& Development, LLC.; Johnson \& Johnson
Pharmaceutical Research \& Development LLC.; Lumosity; Lundbeck; Merck \& Co., Inc.;
Meso Scale Diagnostics, LLC.; NeuroRx Research; Neurotrack Technologies; Novartis
Pharmaceuticals Corporation; Pfizer Inc.; Piramal Imaging; Servier; Takeda Pharmaceutical
Company; and Transition Therapeutics. The Canadian Institutes of Health Research is
providing funds to support ADNI clinical sites in Canada. Private sector contributions are
facilitated by the Foundation for the National Institutes of Health (\href{www.fnih.org}{www.fnih.org}). The grantee
organization is the Northern California Institute for Research and Education, and the study is
coordinated by the Alzheimer’s Therapeutic Research Institute at the University of Southern
California. ADNI data are disseminated by the Laboratory for Neuro Imaging at the
University of Southern California.

Data used in the preparation of this article were obtained from the Alzheimer’s Disease
Neuroimaging Initiative (ADNI) database (\href{adni.loni.usc.edu}{adni.loni.usc.edu}). The ADNI was launched in
2003 as a public-private partnership, led by Principal Investigator Michael W. Weiner,
MD. The primary goal of ADNI has been to test whether serial magnetic resonance imaging
(MRI), positron emission tomography (PET), other biological markers, and clinical and
neuropsychological assessment can be combined to measure the progression of mild
cognitive impairment (MCI) and early Alzheimer’s disease (AD). For up-to-date information,
see \href{www.adni-info.org}{www.adni-info.org}.

\section{Bibliography}

\bibliographystyle{unsrtnat}
\bibliography{ama_ref_2}

\begin{thebibliography}{82}
\providecommand{\natexlab}[1]{#1}
\providecommand{\url}[1]{\texttt{#1}}
\expandafter\ifx\csname urlstyle\endcsname\relax
  \providecommand{\doi}[1]{doi: #1}\else
  \providecommand{\doi}{doi: \begingroup \urlstyle{rm}\Url}\fi

\bibitem[Petersen et~al.(2005)Petersen, Thomas, Grundman, Bennett, Doody,
  Ferris, Galasko, Jin, Kaye, Levey, Pfeiffer, Sano, van Dyck, and
  Thal]{petersen_vitamin_2005}
Ronald~C. Petersen, Ronald~G. Thomas, Michael Grundman, David Bennett, Rachelle
  Doody, Steven Ferris, Douglas Galasko, Shelia Jin, Jeffrey Kaye, Allan Levey,
  Eric Pfeiffer, Mary Sano, Christopher~H. van Dyck, and Leon~J. Thal.
\newblock Vitamin e and donepezil for the treatment of mild cognitive
  impairment.
\newblock \emph{New Engl J Med}, 352\penalty0 (23):\penalty0 2379--2388, June
  2005.
\newblock ISSN 0028-4793, 1533-4406.
\newblock \doi{10.1056/nejmoa050151}.
\newblock URL \url{https://doi.org/10.1056/nejmoa050151}.

\bibitem[Salloway et~al.(2004)Salloway, Ferris, Kluger, Goldman, Griesing,
  Kumar, and Richardson]{salloway_efficacy_2004}
S.~Salloway, S.~Ferris, A.~Kluger, R.~Goldman, T.~Griesing, D.~Kumar, and
  S.~Richardson.
\newblock Efficacy of donepezil in mild cognitive impairment: {A} randomized
  placebo-controlled trial.
\newblock \emph{Neurology}, 63\penalty0 (4):\penalty0 651--657, August 2004.
\newblock ISSN 0028-3878, 1526-632X.
\newblock \doi{10.1212/01.wnl.0000134664.80320.92}.
\newblock URL \url{https://doi.org/10.1212/01.wnl.0000134664.80320.92}.

\bibitem[Doody et~al.(2009)Doody, Ferris, Salloway, Sun, Goldman, Watkins, Xu,
  and Murthy]{doody_donepezil_2009}
R.~S. Doody, S.~H. Ferris, S.~Salloway, Y.~Sun, R.~Goldman, W.~E. Watkins,
  Y.~Xu, and A.~K. Murthy.
\newblock Donepezil treatment of patients with {MCI:} {A} 48-week randomized,
  placebo-controlled trial.
\newblock \emph{Neurology}, 72\penalty0 (18):\penalty0 1555--1561, January
  2009.
\newblock ISSN 0028-3878, 1526-632X.
\newblock \doi{10.1212/01.wnl.0000344650.95823.03}.
\newblock URL \url{https://doi.org/10.1212/01.wnl.0000344650.95823.03}.

\bibitem[Sokolow et~al.(2017)Sokolow, Li, Chen, Taylor, Rotter, Rissman, Aisen,
  and Apostolova]{sokolow_deleterious_2017}
Sophie Sokolow, Xiaohui Li, Lucia Chen, Kent~D. Taylor, Jerome~I. Rotter,
  Robert~A. Rissman, Paul~S. Aisen, and Liana~G. Apostolova.
\newblock Deleterious effect of butyrylcholinesterase k-variant in donepezil
  treatment of mild cognitive impairment.
\newblock \emph{J Alzheimers Dis}, 56\penalty0 (1):\penalty0 229--237, January
  2017.
\newblock ISSN 1387-2877, 1875-8908.
\newblock \doi{10.3233/jad-160562}.
\newblock URL \url{https://doi.org/10.3233/jad-160562}.

\bibitem[Schneider(2011)]{schneider_treatment_2011}
Lon~S. Schneider.
\newblock Treatment with cholinesterase inhibitors and memantine of patients in
  the alzheimer's disease neuroimaging initiative.
\newblock \emph{Arch Neurol}, 68\penalty0 (1):\penalty0 58, January 2011.
\newblock ISSN 0003-9942.
\newblock \doi{10.1001/archneurol.2010.343}.
\newblock URL \url{https://doi.org/10.1001/archneurol.2010.343}.

\bibitem[Devanand et~al.(2018)Devanand, Pelton, D'Antonio, Ciarleglio, Scodes,
  Andrews, Lunsford, Beyer, Petrella, Sneed, Ciovacco, and
  Doraiswamy]{devanand_donepezil_2018}
Davangere~P. Devanand, Gregory~H. Pelton, Kristina D'Antonio, Adam Ciarleglio,
  Jennifer Scodes, Howard Andrews, Julia Lunsford, John~L. Beyer, Jeffrey~R.
  Petrella, Joel Sneed, Michaela Ciovacco, and Pudugramam~Murali Doraiswamy.
\newblock Donepezil treatment in patients with depression and cognitive
  impairment on stable antidepressant treatment: {A} randomized controlled
  trial.
\newblock \emph{Am J Geriatr Psychiatry}, 26\penalty0 (10):\penalty0
  1050--1060, October 2018.
\newblock ISSN 1064-7481.
\newblock \doi{10.1016/j.jagp.2018.05.008}.
\newblock URL \url{https://doi.org/10.1016/j.jagp.2018.05.008}.

\bibitem[Montero-Odasso et~al.(2018)Montero-Odasso, Speechley, Chertkow,
  Sarquis-Adamson, Wells, Borrie, Vanderhaeghe, Zou, Fraser, Bherer, and
  Muir-Hunter]{montero-odasso_donepezil_2019}
M.~Montero-Odasso, M.~Speechley, H.~Chertkow, Y.~Sarquis-Adamson, J.~Wells,
  M.~Borrie, L.~Vanderhaeghe, G.~Y. Zou, S.~Fraser, L.~Bherer, and S.~W.
  Muir-Hunter.
\newblock Donepezil for gait and falls in mild cognitive impairment: {A}
  randomized controlled trial.
\newblock \emph{Eur J Neurol}, 26\penalty0 (4):\penalty0 651--659, December
  2018.
\newblock ISSN 1351-5101, 1468-1331.
\newblock \doi{10.1111/ene.13872}.
\newblock URL \url{https://doi.org/10.1111/ene.13872}.

\bibitem[Petersen et~al.(2010{\natexlab{a}})Petersen, Aisen, Beckett, Donohue,
  Gamst, Harvey, Jack, Jagust, Shaw, Toga, Trojanowski, and
  Weiner]{petersen_alzheimers_2010}
R~C. Petersen, P~S. Aisen, L~A. Beckett, M~C. Donohue, A~C. Gamst, D~J. Harvey,
  C~R. Jack, W~J. Jagust, L~M. Shaw, A~W. Toga, J~Q. Trojanowski, and M~W.
  Weiner.
\newblock Alzheimer's disease neuroimaging initiative ({ADNI}).
\newblock \emph{Neurology}, 74\penalty0 (3):\penalty0 201--209,
  2010{\natexlab{a}}.
\newblock ISSN 0028-3878.
\newblock \doi{10.1212/WNL.0b013e3181cb3e25}.
\newblock URL \url{https://www.ncbi.nlm.nih.gov/pmc/articles/PMC2809036/}.

\bibitem[Baiocchi et~al.(2014)Baiocchi, Cheng, and
  Small]{baiocchi_tutorial_2014}
Michael Baiocchi, Jing Cheng, and Dylan~S. Small.
\newblock Instrumental variable methods for causal inference.
\newblock \emph{Stat Med}, 33\penalty0 (13):\penalty0 2297--2340, March 2014.
\newblock ISSN 0277-6715.
\newblock \doi{10.1002/sim.6128}.
\newblock URL \url{https://doi.org/10.1002/sim.6128}.

\bibitem[van~der Laan(2010)]{van_der_laan_targeted_2010}
Mark~J. van~der Laan.
\newblock Targeted maximum likelihood based causal inference: {Part} i.
\newblock \emph{Int J Biostat}, 6\penalty0 (2):\penalty0 2, January 2010.
\newblock ISSN 1557-4679.
\newblock \doi{10.2202/1557-4679.1211}.
\newblock URL \url{https://doi.org/10.2202/1557-4679.1211}.

\bibitem[Schuler and Rose(2016)]{schuler_targeted_2017}
Megan~S. Schuler and Sherri Rose.
\newblock Targeted maximum likelihood estimation for causal inference in
  observational studies.
\newblock \emph{Am J Epidemiol}, 185\penalty0 (1):\penalty0 65--73, December
  2016.
\newblock ISSN 0002-9262, 1476-6256.
\newblock \doi{10.1093/aje/kww165}.
\newblock URL \url{https://doi.org/10.1093/aje/kww165}.

\bibitem[Chernozhukov et~al.(2018)Chernozhukov, Chetverikov, Demirer, Duflo,
  Hansen, Newey, and Robins]{chernozhukov_doubledebiased_2018}
Victor Chernozhukov, Denis Chetverikov, Mert Demirer, Esther Duflo, Christian
  Hansen, Whitney Newey, and James Robins.
\newblock Double/debiased machine learning for treatment and structural
  parameters.
\newblock \emph{Econom J}, 21\penalty0 (1):\penalty0 C1--C68, January 2018.
\newblock ISSN 1368-4221, 1368-423X.
\newblock \doi{10.1111/ectj.12097}.
\newblock URL \url{https://doi.org/10.1111/ectj.12097}.

\bibitem[Angrist and Frandsen(2022)]{angrist_machine_2022}
Joshua~D. Angrist and Brigham Frandsen.
\newblock Machine labor.
\newblock \emph{J Labor Econ}, 40:\penalty0 S97--S140, 2022.
\newblock ISSN 0734-306X.
\newblock \doi{10.1086/717933}.
\newblock URL \url{https://www.journals.uchicago.edu/doi/abs/10.1086/717933}.
\newblock Publisher: The University of Chicago Press.

\bibitem[Kueper et~al.(2018)Kueper, Speechley, and
  Montero-Odasso]{kueper_alzheimers_2018}
Jacqueline~K. Kueper, Mark Speechley, and Manuel Montero-Odasso.
\newblock The alzheimer{\textquoteright}s disease assessment
  {Scale{\textendash}Cognitive} subscale \{(ADAS\}-cog): {Modifications} and
  responsiveness in pre-dementia populations. a narrative review.
\newblock \emph{J Alzheimers Dis}, 63\penalty0 (2):\penalty0 423--444, April
  2018.
\newblock ISSN 1387-2877, 1875-8908.
\newblock \doi{10.3233/jad-170991}.
\newblock URL \url{https://doi.org/10.3233/jad-170991}.

\bibitem[Rubin(2005)]{rubin_causal_2005}
Donald~B Rubin.
\newblock Causal inference using potential outcomes.
\newblock \emph{J Amer Statist Assoc}, 100\penalty0 (469):\penalty0 322--331,
  March 2005.
\newblock ISSN 0162-1459, 1537-274X.
\newblock \doi{10.1198/016214504000001880}.
\newblock URL \url{https://doi.org/10.1198/016214504000001880}.

\bibitem[Pearl(1995)]{pearl_causal_1995}
Judea Pearl.
\newblock Causal diagrams for empirical research.
\newblock \emph{Biometrika}, 82\penalty0 (4):\penalty0 669--688, December 1995.
\newblock ISSN 0006-3444, 1464-3510.
\newblock \doi{10.1093/biomet/82.4.669}.
\newblock URL \url{https://doi.org/10.1093/biomet/82.4.669}.

\bibitem[VanderWeele and Shpitser(2013)]{vanderweele_definition_2013}
Tyler~J. VanderWeele and Ilya Shpitser.
\newblock On the definition of a confounder.
\newblock \emph{Ann Statist}, 41\penalty0 (1):\penalty0 196--220, February
  2013.
\newblock ISSN 0090-5364.
\newblock \doi{10.1214/12-aos1058}.
\newblock URL \url{https://doi.org/10.1214/12-aos1058}.

\bibitem[Greenland et~al.(1999)Greenland, Pearl, and
  Robins]{greenland_confounding_1999}
Sander Greenland, Judea Pearl, and James~M. Robins.
\newblock Confounding and collapsibility in causal inference.
\newblock \emph{Stat Sci}, 14\penalty0 (1):\penalty0 29--46, February 1999.
\newblock ISSN 0883-4237.
\newblock \doi{10.1214/ss/1009211805}.
\newblock URL \url{https://doi.org/10.1214/ss/1009211805}.

\bibitem[Brookhart et~al.(2010)Brookhart, St\"urmer, Glynn, Rassen, and
  Schneeweiss]{brookhart_confounding_2010}
M.~Alan Brookhart, Til St\"urmer, Robert~J. Glynn, Jeremy Rassen, and Sebastian
  Schneeweiss.
\newblock Confounding control in healthcare database research.
\newblock \emph{Med Care}, 48\penalty0 (6):\penalty0 S114--S120, June 2010.
\newblock ISSN 0025-7079.
\newblock \doi{10.1097/mlr.0b013e3181dbebe3}.
\newblock URL \url{https://doi.org/10.1097/mlr.0b013e3181dbebe3}.

\bibitem[Aronow and Samii(2015)]{aronow_does_2016}
Peter~M. Aronow and Cyrus Samii.
\newblock Does regression produce representative estimates of causal effects?
\newblock \emph{Am J Polit Sci}, 60\penalty0 (1):\penalty0 250--267, April
  2015.
\newblock ISSN 0092-5853, 1540-5907.
\newblock \doi{10.1111/ajps.12185}.
\newblock URL \url{https://doi.org/10.1111/ajps.12185}.

\bibitem[Imbens and Angrist(1994)]{imbens_identification_1994}
Guido~W. Imbens and Joshua~D. Angrist.
\newblock Identification and estimation of local average treatment effects.
\newblock \emph{Econometrica}, 62\penalty0 (2):\penalty0 467, March 1994.
\newblock ISSN 0012-9682.
\newblock \doi{10.2307/2951620}.
\newblock URL \url{https://doi.org/10.2307/2951620}.

\bibitem[Bound et~al.(1995)Bound, Jaeger, and Baker]{bound_problems_1995}
John Bound, David~A. Jaeger, and Regina~M. Baker.
\newblock Problems with instrumental variables estimation when the correlation
  between the instruments and the endogeneous explanatory variable is weak.
\newblock \emph{J Amer Statist Assoc}, 90\penalty0 (430):\penalty0 443, June
  1995.
\newblock ISSN 0162-1459.
\newblock \doi{10.2307/2291055}.
\newblock URL \url{https://doi.org/10.2307/2291055}.

\bibitem[Wooldridge(2010)]{wooldridge_econometric_2010}
Jeffrey~M. Wooldridge.
\newblock \emph{Econometric Analysis of Cross Section and Panel Data}.
\newblock MIT Press, Cambridge, MA, USA, 2 edition, October 2010.
\newblock ISBN 978-0-262-23258-6.

\bibitem[Angrist and Pischke(2009)]{angrist_mostly_2009}
Joshua~D Angrist and Jorn-Steffen Pischke.
\newblock \emph{Mostly {Harmless} {\{Econometrics\}:} {An}
  {\{Empiricist\}{\textquoteright}s} {Companion}}.
\newblock Princeton University Press, 1 edition, January 2009.

\bibitem[Abadie(2003)]{abadie_semiparametric_2003}
Alberto Abadie.
\newblock Semiparametric instrumental variable estimation of treatment response
  models.
\newblock \emph{J Econometrics}, 113\penalty0 (2):\penalty0 231--263, April
  2003.
\newblock ISSN 0304-4076.
\newblock \doi{10.1016/s0304-4076(02)00201-4}.
\newblock URL \url{https://doi.org/10.1016/s0304-4076(02)00201-4}.

\bibitem[Blandhol et~al.(2022)Blandhol, Bonney, Mogstad, and
  Torgovitsky]{blandhol_when_2022}
Christine Blandhol, John Bonney, Magne Mogstad, and Alexander Torgovitsky.
\newblock When is tsls actually late?
\newblock \emph{SSRN Electronic Journal}, page~69, February 2022.
\newblock ISSN 1556-5068.
\newblock \doi{10.2139/ssrn.4021804}.
\newblock URL \url{https://doi.org/10.2139/ssrn.4021804}.

\bibitem[Aronow and Carnegie(2017)]{aronow_beyond_2017}
Peter~M. Aronow and Allison Carnegie.
\newblock Beyond {LATE}: Estimation of the average treatment effect with an
  instrumental variable.
\newblock \emph{Polit Anal}, 21\penalty0 (4):\penalty0 492--506, 2017.
\newblock ISSN 1047-1987, 1476-4989.
\newblock \doi{10.1093/pan/mpt013}.
\newblock URL
  \url{https://www.cambridge.org/core/journals/political-analysis/article/abs/beyond-late-estimation-of-the-average-treatment-effect-with-an-instrumental-variable/604E0803793175CF88329DB34DAA80B3}.

\bibitem[Wang and Tchetgen~Tchetgen(2020)]{wang_bounded_2018}
Linbo Wang and Eric Tchetgen~Tchetgen.
\newblock Bounded, efficient and multiply robust estimation of average
  treatment effects using instrumental variables.
\newblock \emph{J R Stat Soc Series B Stat Methodol}, 80\penalty0 (3):\penalty0
  531--550, 2020.
\newblock ISSN 1369-7412.
\newblock \doi{10.1111/rssb.12262}.

\bibitem[Hartwig et~al.(2020)Hartwig, Wang, Smith, and
  Davies]{hartwig_average_2020}
F.~P. Hartwig, L.~Wang, G.~Davey Smith, and N.~M. Davies.
\newblock Average causal effect estimation via instrumental variables: the no
  simultaneous heterogeneity assumption.
\newblock \emph{arXiv}, 2020.
\newblock \doi{10.48550/arXiv.2010.10017}.
\newblock URL \url{https://arxiv.org/abs/2010.10017v3}.

\bibitem[Wennberg(1984)]{wennberg_dealing_1984}
John~E. Wennberg.
\newblock Dealing with medical practice variations: {A} proposal for action.
\newblock \emph{Health Affair}, 3\penalty0 (2):\penalty0 6--33, January 1984.
\newblock ISSN 0278-2715, 1544-5208.
\newblock \doi{10.1377/hlthaff.3.2.6}.
\newblock URL \url{https://doi.org/10.1377/hlthaff.3.2.6}.

\bibitem[Wennberg and Gittelsohn(1973)]{wennberg_small_1973}
John Wennberg and Alan Gittelsohn.
\newblock Small area variations in health care delivery.
\newblock \emph{Science}, 182\penalty0 (4117):\penalty0 1102--1108, December
  1973.
\newblock ISSN 0036-8075, 1095-9203.
\newblock \doi{10.1126/science.182.4117.1102}.
\newblock URL \url{https://doi.org/10.1126/science.182.4117.1102}.

\bibitem[Corallo et~al.(2014)Corallo, Croxford, Goodman, Bryan, Srivastava, and
  Stukel]{corallo_systematic_2014}
Ashley~N. Corallo, Ruth Croxford, David~C. Goodman, Elisabeth~L. Bryan, Divya
  Srivastava, and Therese~A. Stukel.
\newblock A systematic review of medical practice variation in {OECD}
  countries.
\newblock \emph{Health Policy}, 114\penalty0 (1):\penalty0 5--14, January 2014.
\newblock ISSN 0168-8510.
\newblock \doi{10.1016/j.healthpol.2013.08.002}.
\newblock URL \url{https://doi.org/10.1016/j.healthpol.2013.08.002}.

\bibitem[Chen and Briesacher(2011)]{chen_use_2011}
Yong Chen and Becky~A. Briesacher.
\newblock Use of instrumental variable in prescription drug research with
  observational data: {A} systematic review.
\newblock \emph{J Clin Epidemiol}, 64\penalty0 (6):\penalty0 687--700, June
  2011.
\newblock ISSN 0895-4356.
\newblock \doi{10.1016/j.jclinepi.2010.09.006}.
\newblock URL \url{https://doi.org/10.1016/j.jclinepi.2010.09.006}.

\bibitem[Brookhart and Schneeweiss(2007)]{brookhart_preference-based_2007}
M.~Alan Brookhart and Sebastian Schneeweiss.
\newblock Preference-based instrumental variable methods for the estimation of
  treatment effects: {Assessing} validity and interpreting results.
\newblock \emph{Int J Biostat}, 3\penalty0 (1):\penalty0 14, January 2007.
\newblock ISSN 1557-4679.
\newblock \doi{10.2202/1557-4679.1072}.
\newblock URL \url{https://doi.org/10.2202/1557-4679.1072}.

\bibitem[Hirano and Imbens(2001)]{hirano_estimation_2001}
Keisuke Hirano and Guido~W. Imbens.
\newblock Estimation of causal effects using propensity score weighting: {An}
  application to data on right heart catheterization.
\newblock \emph{Health Serv Outcomes Res Methodol}, 2\penalty0 (3/4):\penalty0
  259--278, December 2001.
\newblock ISSN 1387-3741.
\newblock \doi{10.1023/a:1020371312283}.
\newblock URL \url{https://doi.org/10.1023/a:1020371312283}.

\bibitem[Bhattacharya and Vogt(2007)]{bhattacharya_instrumental_2007}
Jay Bhattacharya and William Vogt.
\newblock Do instrumental variables belong in propensity scores?
\newblock Technical Report t0343, National Bureau of Economic Research,
  Cambridge, MA, September 2007.
\newblock URL \url{https://doi.org/10.3386/t0343}.

\bibitem[Wooldridge(2016)]{wooldridge_should_2016}
Jeffrey~M. Wooldridge.
\newblock Should instrumental variables be used as matching variables?
\newblock \emph{Research in Economics}, 70\penalty0 (2):\penalty0 232--237,
  June 2016.
\newblock ISSN 1090-9443.
\newblock \doi{10.1016/j.rie.2016.01.001}.
\newblock URL
  \url{https://www.sciencedirect.com/science/article/pii/S1090944315301678}.

\bibitem[Ding et~al.(2017)Ding, Vanderweele, and
  Robins]{ding_instrumental_2017}
P.~Ding, T.J. Vanderweele, and J.~M. Robins.
\newblock Instrumental variables as bias amplifiers with general outcome and
  confounding.
\newblock \emph{Biometrika}, 104\penalty0 (2):\penalty0 291--302, April 2017.
\newblock ISSN 0006-3444, 1464-3510.
\newblock \doi{10.1093/biomet/asx009}.
\newblock URL \url{https://doi.org/10.1093/biomet/asx009}.

\bibitem[Pearl(2010)]{pearl_class_2010}
Judea Pearl.
\newblock On a class of bias-amplifying variables that endanger effect
  estimates, 2010.

\bibitem[Brookhart et~al.(2006)Brookhart, Schneeweiss, Rothman, Glynn, Avorn,
  and St\"urmer]{brookhart_variable_2006}
M.~Alan Brookhart, Sebastian Schneeweiss, Kenneth~J. Rothman, Robert~J. Glynn,
  Jerry Avorn, and Til St\"urmer.
\newblock Variable selection for propensity score models.
\newblock \emph{Am J Epidemiol}, 163\penalty0 (12):\penalty0 1149--1156, April
  2006.
\newblock ISSN 1476-6256, 0002-9262.
\newblock \doi{10.1093/aje/kwj149}.
\newblock URL \url{https://doi.org/10.1093/aje/kwj149}.

\bibitem[Brooks and Ohsfeldt(2012)]{brooks_squeezing_2013}
John~M. Brooks and Robert~L. Ohsfeldt.
\newblock Squeezing the balloon: {Propensity} scores and unmeasured covariate
  balance.
\newblock \emph{Health Serv Res}, 48\penalty0 (4):\penalty0 1487--1507,
  December 2012.
\newblock ISSN 0017-9124.
\newblock \doi{10.1111/1475-6773.12020}.
\newblock URL \url{https://doi.org/10.1111/1475-6773.12020}.

\bibitem[Petersen et~al.(2010{\natexlab{b}})Petersen, Porter, Gruber, Wang, and
  van~der Laan]{petersen_diagnosing_2012}
Maya~L Petersen, Kristin~E Porter, Susan Gruber, Yue Wang, and Mark~J van~der
  Laan.
\newblock Diagnosing and responding to violations in the positivity assumption.
\newblock \emph{Stat Methods Med Res}, 21\penalty0 (1):\penalty0 31--54,
  October 2010{\natexlab{b}}.
\newblock ISSN 0962-2802, 1477-0334.
\newblock \doi{10.1177/0962280210386207}.
\newblock URL \url{https://doi.org/10.1177/0962280210386207}.

\bibitem[Westreich and Cole(2010)]{westreich_invited_2010}
D.~Westreich and S.~R. Cole.
\newblock Invited commentary: {Positivity} in practice.
\newblock \emph{Am J Epidemiol}, 171\penalty0 (6):\penalty0 674--677, February
  2010.
\newblock ISSN 0002-9262, 1476-6256.
\newblock \doi{10.1093/aje/kwp436}.
\newblock URL \url{https://doi.org/10.1093/aje/kwp436}.

\bibitem[Robins et~al.(2000)Robins, Hern\'an, and
  Brumback]{robins_marginal_2000}
James~M. Robins, Miguel~\'Angel Hern\'an, and Babette Brumback.
\newblock Marginal structural models and causal inference in epidemiology.
\newblock \emph{Epidemiology}, 11\penalty0 (5):\penalty0 550--560, September
  2000.
\newblock ISSN 1044-3983.
\newblock \doi{10.1097/00001648-200009000-00011}.
\newblock URL \url{https://doi.org/10.1097/00001648-200009000-00011}.

\bibitem[Rosenbaum and Rubin(1983)]{rosenbaum_central_1983}
Paul~R. Rosenbaum and Donald~B. Rubin.
\newblock The central role of the propensity score in observational studies for
  causal effects.
\newblock \emph{Biometrika}, 70\penalty0 (1):\penalty0 41--55, April 1983.
\newblock ISSN 0006-3444, 1464-3510.
\newblock \doi{10.1093/biomet/70.1.41}.
\newblock URL \url{https://doi.org/10.1093/biomet/70.1.41}.

\bibitem[Austin and Stuart(2015)]{austin_moving_2015}
Peter~C. Austin and Elizabeth~A. Stuart.
\newblock Moving towards best practice when using inverse probability of
  treatment weighting {(IPTW)} using the propensity score to estimate causal
  treatment effects in observational studies.
\newblock \emph{Stat Med}, 34\penalty0 (28):\penalty0 3661--3679, August 2015.
\newblock ISSN 0277-6715, 1097-0258.
\newblock \doi{10.1002/sim.6607}.
\newblock URL \url{https://doi.org/10.1002/sim.6607}.

\bibitem[Vegetabile et~al.(2020)Vegetabile, Gillen, and
  Stern]{vegetabile_optimally_2020}
Brian~G. Vegetabile, Daniel~L. Gillen, and Hal~S. Stern.
\newblock Optimally balanced gaussian process propensity scores for estimating
  treatment effects.
\newblock \emph{J R Stat Soc Ser A Stat Soc}, 183\penalty0 (1):\penalty0
  355--377, 2020.
\newblock ISSN 0964-1998.
\newblock \doi{10.1111/rssa.12502}.
\newblock URL \url{https://www.ncbi.nlm.nih.gov/pmc/articles/PMC8360444/}.

\bibitem[Imai and Ratkovic(2014)]{imai_covariate_2014}
Kosuke Imai and Marc Ratkovic.
\newblock Covariate balancing propensity score.
\newblock \emph{J R Stat Soc Ser B Methodol}, 76\penalty0 (1):\penalty0
  243--263, 2014.
\newblock ISSN 1467-9868.
\newblock \doi{10.1111/rssb.12027}.
\newblock URL \url{https://onlinelibrary.wiley.com/doi/abs/10.1111/rssb.12027}.

\bibitem[Weitzen et~al.(2004)Weitzen, Lapane, Toledano, Hume, and
  Mor]{weitzen_weaknesses_2005}
Sherry Weitzen, Kate~L. Lapane, Alicia~Y. Toledano, Anne~L. Hume, and Vincent
  Mor.
\newblock Weaknesses of goodness-of-fit tests for evaluating propensity score
  models: {The} case of the omitted confounder.
\newblock \emph{Pharmacoepidem Dr S}, 14\penalty0 (4):\penalty0 227--238, July
  2004.
\newblock ISSN 1053-8569.
\newblock \doi{10.1002/pds.986}.
\newblock URL \url{https://doi.org/10.1002/pds.986}.

\bibitem[Chernozhukov et~al.(2015)Chernozhukov, Hansen, and
  Spindler]{chernozhukov_post-selection_2015}
Victor Chernozhukov, Christian Hansen, and Martin Spindler.
\newblock Post-selection and post-regularization inference in linear models
  with many controls and instruments.
\newblock \emph{Am. Econ. Rev.}, 105\penalty0 (5):\penalty0 486--490, May 2015.
\newblock ISSN 0002-8282.
\newblock \doi{10.1257/aer.p20151022}.
\newblock URL \url{https://doi.org/10.1257/aer.p20151022}.

\bibitem[Kinal(1980)]{kinal_existence_1980}
Terrence~W. Kinal.
\newblock The existence of moments of k-class estimators.
\newblock \emph{Econometrica}, 48\penalty0 (1):\penalty0 241, January 1980.
\newblock ISSN 0012-9682.
\newblock \doi{10.2307/1912027}.
\newblock URL \url{https://doi.org/10.2307/1912027}.

\bibitem[Angrist and Krueger(1995)]{angrist_split-sample_1995}
Joshua~D. Angrist and Alan~B. Krueger.
\newblock Split-sample instrumental variables estimates of the return to
  schooling.
\newblock \emph{Journal of Business \&amp; Economic Statistics}, 13\penalty0
  (2):\penalty0 225--235, April 1995.
\newblock ISSN 0735-0015, 1537-2707.
\newblock \doi{10.1080/07350015.1995.10524597}.
\newblock URL \url{https://doi.org/10.1080/07350015.1995.10524597}.

\bibitem[Roodman(2009)]{roodman_note_2009}
David Roodman.
\newblock A note on the theme of too many instruments.
\newblock \emph{Oxford B Econ Stat}, 71\penalty0 (1):\penalty0 135--158,
  February 2009.
\newblock ISSN 0305-9049, 1468-0084.
\newblock \doi{10.1111/j.1468-0084.2008.00542.x}.
\newblock URL \url{https://doi.org/10.1111/j.1468-0084.2008.00542.x}.

\bibitem[Young(2022)]{young_consistency_2022}
Alwyn Young.
\newblock Consistency without {Inference}: {Instrumental} {Variables} in
  {Practical} {Application}.
\newblock \emph{Eur Econ Rev}, 147:\penalty0 104112, August 2022.
\newblock ISSN 0014-2921.
\newblock \doi{10.1016/j.euroecorev.2022.104112}.
\newblock URL
  \url{https://www.sciencedirect.com/science/article/pii/S001429212200054X}.

\bibitem[Guggenberger(2011)]{guggenberger_asymptotic_2012}
Patrik Guggenberger.
\newblock On the asymptotic size distortion of tests when instruments locally
  violate the exogeneity assumption.
\newblock \emph{Economet Theor}, 28\penalty0 (2):\penalty0 387--421, September
  2011.
\newblock ISSN 0266-4666, 1469-4360.
\newblock \doi{10.1017/s0266466611000375}.
\newblock URL \url{https://doi.org/10.1017/s0266466611000375}.

\bibitem[Davies et~al.(2017)Davies, Thomas, Taylor, Taylor, Martin, Munaf\`o,
  and Windmeijer]{davies_how_2017}
Neil~M Davies, Kyla~H Thomas, Amy~E Taylor, Gemma~MJ Taylor, Richard~M Martin,
  Marcus~R Munaf\`o, and Frank Windmeijer.
\newblock How to compare instrumental variable and conventional regression
  analyses using negative controls and bias plots.
\newblock \emph{Int. J. Epidemiol.}, 46\penalty0 (6):\penalty0 2067--2077,
  April 2017.
\newblock ISSN 0300-5771, 1464-3685.
\newblock \doi{10.1093/ije/dyx014}.
\newblock URL \url{https://doi.org/10.1093/ije/dyx014}.

\bibitem[Sargan(1958)]{Sargan_estimation_1958}
J.~D. Sargan.
\newblock The estimation of economic relationships using instrumental
  variables.
\newblock \emph{Econometrica}, 26\penalty0 (3):\penalty0 393, July 1958.
\newblock ISSN 0012-9682.
\newblock \doi{10.2307/1907619}.
\newblock URL \url{https://doi.org/10.2307/1907619}.

\bibitem[Kiviet and Kripfganz(2021)]{kiviet_instrument_2021}
Jan~F. Kiviet and Sebastian Kripfganz.
\newblock Instrument approval by the sargan test and its consequences for
  coefficient estimation.
\newblock \emph{Econ Lett}, 205:\penalty0 109935, August 2021.
\newblock ISSN 0165-1765.
\newblock \doi{10.1016/j.econlet.2021.109935}.
\newblock URL \url{https://doi.org/10.1016/j.econlet.2021.109935}.

\bibitem[Pearl(2009)]{pearl_causality_2009}
Judea Pearl.
\newblock \emph{Causality}.
\newblock Cambridge University Press, Cambridge, 2 edition, September 2009.
\newblock ISBN 9780511803161, 9780521895606, 9780521749190.
\newblock \doi{10.1017/cbo9780511803161}.
\newblock URL \url{https://doi.org/10.1017/cbo9780511803161}.

\bibitem[Mood(2009)]{mood_logistic_2010}
C.~Mood.
\newblock Logistic regression: {Why} we cannot do what we think we can do, and
  what we can do about it.
\newblock \emph{Eur Sociol Rev}, 26\penalty0 (1):\penalty0 67--82, March 2009.
\newblock ISSN 0266-7215, 1468-2672.
\newblock \doi{10.1093/esr/jcp006}.
\newblock URL \url{https://doi.org/10.1093/esr/jcp006}.

\bibitem[Schuster et~al.(2021)Schuster, Twisk, ter Riet, Heymans, and
  Rijnhart]{schuster_noncollapsibility_2021}
Noah~A. Schuster, Jos W.~R. Twisk, Gerben ter Riet, Martijn~W. Heymans, and
  Judith J.~M. Rijnhart.
\newblock Noncollapsibility and its role in quantifying confounding bias in
  logistic regression.
\newblock \emph{BMC Med Res Methodol}, 21\penalty0 (1):\penalty0 136, July
  2021.
\newblock ISSN 1471-2288.
\newblock \doi{10.1186/s12874-021-01316-8}.
\newblock URL \url{https://doi.org/10.1186/s12874-021-01316-8}.

\bibitem[Schisterman et~al.(2009)Schisterman, Cole, and
  Platt]{schisterman_overadjustment_2009}
Enrique~F. Schisterman, Stephen~R. Cole, and Robert~W. Platt.
\newblock Overadjustment bias and unnecessary adjustment in epidemiologic
  studies.
\newblock \emph{Epidemiology}, 20\penalty0 (4):\penalty0 488--495, 2009.
\newblock ISSN 1044-3983.
\newblock \doi{10.1097/EDE.0b013e3181a819a1}.
\newblock URL
  \url{https://journals.lww.com/epidem/Fulltext/2009/07000/Overadjustment_Bias_and_Unnecessary_Adjustment_in.4.aspx}.

\bibitem[Karlson et~al.(2021)Karlson, Popham, and Holm]{karlson_marginal_2021}
Kristian~Bernt Karlson, Frank Popham, and Anders Holm.
\newblock Marginal and conditional confounding using logits.
\newblock \emph{Sociol Methods Res}, page 004912412199554, April 2021.
\newblock ISSN 0049-1241, 1552-8294.
\newblock \doi{10.1177/0049124121995548}.
\newblock URL \url{https://doi.org/10.1177/0049124121995548}.

\bibitem[Vansteelandt and Keiding(2011)]{vansteelandt_invited_2011}
S.~Vansteelandt and N.~Keiding.
\newblock Invited commentary: {G-computation-lost} in translation?
\newblock \emph{Am J Epidemiol}, 173\penalty0 (7):\penalty0 739--742, March
  2011.
\newblock ISSN 0002-9262, 1476-6256.
\newblock \doi{10.1093/aje/kwq474}.
\newblock URL \url{https://doi.org/10.1093/aje/kwq474}.

\bibitem[Pearl and Paz(2014)]{pearl_confounding_2014}
Judea Pearl and Azaria Paz.
\newblock Confounding equivalence in causal inference.
\newblock \emph{J Causal Inference}, 2\penalty0 (1):\penalty0 75--93, March
  2014.
\newblock ISSN 2193-3677, 2193-3685.
\newblock \doi{10.1515/jci-2013-0020}.
\newblock URL \url{https://doi.org/10.1515/jci-2013-0020}.

\bibitem[Basu et~al.(2018)Basu, Coe, and Chapman]{basu_2sls_2018}
Anirban Basu, Norma~B. Coe, and Cole~G. Chapman.
\newblock {2SLS} versus {2SRI:} {Appropriate} methods for rare outcomes and/or
  rare exposures.
\newblock \emph{Health Econ}, 27\penalty0 (6):\penalty0 937--955, March 2018.
\newblock ISSN 1057-9230.
\newblock \doi{10.1002/hec.3647}.
\newblock URL \url{https://doi.org/10.1002/hec.3647}.

\bibitem[Chapman and Brooks(2016)]{chapman_treatment_2016}
Cole~G. Chapman and John~M. Brooks.
\newblock Treatment effect estimation using nonlinear two-stage instrumental
  variable estimators: {Another} cautionary note.
\newblock \emph{Health Serv Res}, 51\penalty0 (6):\penalty0 2375--2394,
  February 2016.
\newblock ISSN 0017-9124.
\newblock \doi{10.1111/1475-6773.12463}.
\newblock URL \url{https://doi.org/10.1111/1475-6773.12463}.

\bibitem[Horrace and Oaxaca(2006)]{horrace_results_2006}
William~C. Horrace and Ronald~L. Oaxaca.
\newblock Results on the bias and inconsistency of ordinary least squares for
  the linear probability model.
\newblock \emph{Econ Lett}, 90\penalty0 (3):\penalty0 321--327, March 2006.
\newblock ISSN 0165-1765.
\newblock \doi{10.1016/j.econlet.2005.08.024}.
\newblock URL \url{https://doi.org/10.1016/j.econlet.2005.08.024}.

\bibitem[Belloni et~al.(2012)Belloni, Chen, Chernozhukov, and
  Hansen]{belloni_sparse_2012}
A.~Belloni, D.~Chen, V.~Chernozhukov, and C.~Hansen.
\newblock Sparse models and methods for optimal instruments with an application
  to eminent domain.
\newblock \emph{Econometrica}, 80\penalty0 (6):\penalty0 2369--2429, 2012.
\newblock ISSN 0012-9682.
\newblock \doi{10.3982/ecta9626}.
\newblock URL \url{https://doi.org/10.3982/ecta9626}.

\bibitem[Luque-Fernandez et~al.(2018)Luque-Fernandez, Schomaker, Rachet, and
  Schnitzer]{luquefernandez_targeted_2018}
Miguel~Angel Luque-Fernandez, Michael Schomaker, Bernard Rachet, and
  Mireille~E. Schnitzer.
\newblock Targeted maximum likelihood estimation for a binary treatment: {A}
  tutorial.
\newblock \emph{Stat Med}, 37\penalty0 (16):\penalty0 2530--2546, April 2018.
\newblock ISSN 0277-6715.
\newblock \doi{10.1002/sim.7628}.
\newblock URL \url{https://doi.org/10.1002/sim.7628}.

\bibitem[D{\'\i}az(2019)]{diaz_machine_2020}
Iv\'an D{\'\i}az.
\newblock Machine learning in the estimation of causal effects: {Targeted}
  minimum loss-based estimation and double/debiased machine learning.
\newblock \emph{Biostatistics}, 21\penalty0 (2):\penalty0 353--358, November
  2019.
\newblock ISSN 1465-4644, 1468-4357.
\newblock \doi{10.1093/biostatistics/kxz042}.
\newblock URL \url{https://doi.org/10.1093/biostatistics/kxz042}.

\bibitem[van~der Laan and Robins(2003)]{van_der_laan_unified_2003}
Mark~J. van~der Laan and James~M. Robins.
\newblock \emph{Unified Methods for Censored Longitudinal Data and Causality}.
\newblock Springer Series in Statistics. Springer, 2003.
\newblock ISBN 978-1-4419-3055-2 978-0-387-21700-0.
\newblock \doi{10.1007/978-0-387-21700-0}.
\newblock URL \url{http://link.springer.com/10.1007/978-0-387-21700-0}.

\bibitem[Neyman(1959)]{neyman_jerzy_optimal_1959}
Jerzy Neyman.
\newblock Optimal asymptotic tests of composite statistical hypotheses.
\newblock \emph{Probability and Statistics}, pages 416--444, 1959.

\bibitem[van~der Ploeg et~al.(2014)van~der Ploeg, Austin, and
  Steyerberg]{van_der_ploeg_modern_2014}
Tjeerd van~der Ploeg, Peter~C Austin, and Ewout~W Steyerberg.
\newblock Modern modelling techniques are data hungry: {A} simulation study for
  predicting dichotomous endpoints.
\newblock \emph{BMC Med Res Methodol}, 14\penalty0 (1):\penalty0 137, December
  2014.
\newblock ISSN 1471-2288.
\newblock \doi{10.1186/1471-2288-14-137}.
\newblock URL \url{https://doi.org/10.1186/1471-2288-14-137}.

\bibitem[Andrews et~al.(2019)Andrews, Stock, and Sun]{andrews_weak_2019}
Isaiah Andrews, James~H. Stock, and Liyang Sun.
\newblock Weak instruments in instrumental variables regression: {Theory} and
  practice.
\newblock \emph{Annu Rev Econ}, 11\penalty0 (1):\penalty0 727--753, August
  2019.
\newblock ISSN 1941-1383, 1941-1391.
\newblock \doi{10.1146/annurev-economics-080218-025643}.
\newblock URL \url{https://doi.org/10.1146/annurev-economics-080218-025643}.

\bibitem[Windmeijer et~al.(2018)Windmeijer, Farbmacher, Davies, and
  Davey~Smith]{windmeijer_use_2018}
Frank Windmeijer, Helmut Farbmacher, Neil Davies, and George Davey~Smith.
\newblock On the use of the lasso for instrumental variables estimation with
  some invalid instruments.
\newblock \emph{J Amer Statist Assoc}, 114\penalty0 (527):\penalty0 1339--1350,
  November 2018.
\newblock ISSN 0162-1459, 1537-274X.
\newblock \doi{10.1080/01621459.2018.1498346}.
\newblock URL \url{https://doi.org/10.1080/01621459.2018.1498346}.

\bibitem[VanderWeele(2012)]{vander_weele_confounding_2012}
Tyler~J. VanderWeele.
\newblock Confounding and effect modification: {Distribution} and measure.
\newblock \emph{Epidemiol Methods}, 1\penalty0 (1):\penalty0 55--82, January
  2012.
\newblock ISSN 2161-962X.
\newblock \doi{10.1515/2161-962x.1004}.
\newblock URL \url{https://doi.org/10.1515/2161-962x.1004}.

\bibitem[Jackson and Swanson(2015)]{jackson_toward_2015}
John~W. Jackson and Sonja~A. Swanson.
\newblock Toward a clearer portrayal of confounding bias in instrumental
  variable applications.
\newblock \emph{Epidemiology}, 26\penalty0 (4):\penalty0 498--504, July 2015.
\newblock ISSN 1044-3983.
\newblock \doi{10.1097/ede.0000000000000287}.
\newblock URL \url{https://doi.org/10.1097/ede.0000000000000287}.

\bibitem[Zhao and Small(2018)]{zhao_graphical_2018}
Qingyuan Zhao and Dylan~S. Small.
\newblock Graphical diagnosis of confounding bias in instrumental variable
  analysis.
\newblock \emph{Epidemiology}, 29\penalty0 (4):\penalty0 e29--e31, July 2018.
\newblock ISSN 1044-3983.
\newblock \doi{10.1097/ede.0000000000000822}.
\newblock URL \url{https://doi.org/10.1097/ede.0000000000000822}.

\bibitem[Mueller et~al.(2005)Mueller, Weiner, Thal, Petersen, Jack, Jagust,
  Trojanowski, Toga, and Beckett]{mueller_ways_2005}
Susanne~G. Mueller, Michael~W. Weiner, Leon~J. Thal, Ronald~C. Petersen,
  Clifford~R. Jack, William Jagust, John~Q. Trojanowski, Arthur~W. Toga, and
  Laurel Beckett.
\newblock Ways toward an early diagnosis in alzheimer's disease: {The}
  alzheimer's disease neuroimaging initiative {(ADNI)}.
\newblock \emph{Alzheimers Dement}, 1\penalty0 (1):\penalty0 55--66, July 2005.
\newblock ISSN 1552-5260, 1552-5279.
\newblock \doi{10.1016/j.jalz.2005.06.003}.
\newblock URL \url{https://doi.org/10.1016/j.jalz.2005.06.003}.

\bibitem[Gupta(2011)]{gupta_intention--treat_2011}
SandeepK Gupta.
\newblock Intention-to-treat concept: {A} review.
\newblock \emph{Perspect Clin Res}, 2\penalty0 (3):\penalty0 109, 2011.
\newblock ISSN 2229-3485.
\newblock \doi{10.4103/2229-3485.83221}.
\newblock URL \url{https://doi.org/10.4103/2229-3485.83221}.

\bibitem[Chernozhukov et~al.(2016)Chernozhukov, Hansen, and
  Spindler]{chernozhukov_high-dimensional_2016}
Victor Chernozhukov, Chris Hansen, and Martin Spindler.
\newblock High-{Dimensional} {Metrics} in {R}.
\newblock \emph{arXiv}, August 2016.
\newblock \doi{10.48550/arXiv.1603.01700}.
\newblock URL \url{http://arxiv.org/abs/1603.01700}.

\end{thebibliography}

\end{document}